\documentclass[letterpaper]{JHEP3}

\usepackage{amsmath,amsthm, amssymb}

\newcommand{\be}{\begin{equation}}
\newcommand{\ee}{\end{equation}}
\newcommand{\ba}{\begin{array}}
\newcommand{\ea}{\end{array}}

\newcommand{\bea}{\begin{eqnarray}}
\newcommand{\eea}{\end{eqnarray}}
\newcommand{\bma}{\begin{matrix}}
\newcommand{\ema}{\end{matrix}}
\newcommand{\bpm}{\begin{pmatrix}}
\newcommand{\epm}{\end{pmatrix}}
\newcommand{\nn}{\nonumber}
\newcommand{\mc}{\mathcal}

\newcommand{\qq}{\vec q ^{\,2}}
\newcommand{\bvec}[1]{\stackrel{\leftarrow}{#1}}
\newcommand{\dvec}[1]{\stackrel{\rightarrow}{#1}}

\newcommand{\sx}{\sqrt6}

\newcommand{\p}{\partial}
\newcommand{\hf}{{\hat5}}
\newcommand{\Qij}{Q_i{}^j}
\newcommand{\Uij}{U_i{}^j}

\newcommand{\ov}{\overline}
\newcommand{\wh}{\widehat}
\newcommand{\wt}{\widetilde}
\newcommand{\Psitil}{\wt \Psi}
\newcommand{\Psibar}{\ov \Psi}
\newcommand{\Zeta}{\Upsilon}
\newcommand{\Eta}{\mathcal H}
\newcommand{\Etatil}{\wt \Eta}
\newcommand{\psibar}{\ov \psi}
\newcommand{\etabar}{\ov \eta}
\newcommand{\sibar}{\ov \sigma}

\newcommand{\xibar}{\ov \xi}

\newcommand{\ep}{\varepsilon}
\newcommand{\eps}{\epsilon}
\newcommand{\al}{\alpha}
\newcommand{\la}{\lambda}
\newcommand{\da}{\delta}
\newcommand{\om}{\omega}
\newcommand{\Ga}{\Gamma}
\newcommand{\ga}{\gamma}

\newcommand{\Si}{\Sigma}
\newcommand{\si}{\sigma}

\title{\bf Boundary conditions in supergravity
on a manifold with boundary}
\author{Dmitry V. Belyaev\footnote{Present address:
DESY-T, Notkestrasse 85, 22603 Hamburg, Germany}
\\
Department of Physics and Astronomy, The Johns Hopkins University, \\
3400 North Charles Street, Baltimore, MD 21218, USA\\ 
E-mail: \email{dmitry.belyaev@desy.de}}

\preprint{}

\abstract{
We explain why it is necessary to use boundary conditions 
in the proof of supersymmetry of a supergravity action on
a manifold with boundary. Working in both boundary (``downstairs'')
and orbifold (``upstairs'') pictures, we present a
bulk-plus-boundary/brane action for the five-dimensional
(on-shell) supergravity which is supersymmetric with the
use of \emph{fewer} boundary conditions than 
were previously employed. The required Gibbons-Hawking-like 
$Y$-term and many other aspects of the boundary/orbifold picture
correspondence are discussed.
}

\begin{document}
\section{Introduction}
In their ground-breaking paper \cite{hw}, Horava and Witten
discussed eleven-dimensional supergravity on a manifold with boundary, 
that arises as a low energy limit of the strongly coupled
heterotic string theory. 
They explained that there are two possible descriptions of the same
theory: the ``downstairs'' (boundary) picture and 
the ``upstairs'' (orbifold) picture. They made a comment, however, 
that working on the orbifold is technically more convenient,
which was, perhaps, the reason why the orbifold picture became
the de facto choice for many researchers working in this area.

One of such orbifold constructions is the now famous Randall-Sundrum
scenario \cite{rs1,rs2} which is set up in a five-dimensional
space-time with a negative cosmological constant. This scenario
was supersymmetrized by different groups \cite{abn, gp1, flp1} 
with somewhat different approaches
(see Ref.~\cite{bb1} for the proof of the equivalence of these
approaches). The original orbifold construction was used for
the supersymmetrization as well.

Over time people came to discover that the boundary 
(``downstairs'' or ``interval'') picture is in many respects
preferred over the orbifold picture. For example, in Ref.~\cite{bb2}
it was demonstrated how the same physical content unambiguously
encoded in the boundary picture can be obscured by various ``twists''
and ``jumps'' on the orbifold.

In this paper we provide more evidence for simplicity of 
the boundary picture.

We present a bulk-plus-boundary action with
the five-dimensional gauged (on-shell) supergravity in the bulk.
It is $N=2$ (locally) supersymmetric with a boundary condition on the 
supersymmetry parameter breaking a half of the bulk supersymmetries
on the boundary.
Supersymmetry of the action requires the use of a 
\emph{small subset} of all the (natural) boundary conditions
encoded in the action itself. 
Which boundary conditions are necessary is indicated by the
supersymmetry algebra.

The boundary action is a sum of two terms. The first one
is a Gibbons-Hawking-like term \cite{gibh} (which we call ``$Y$-term''
to acknowledge the work of York \cite{york1, york2}).
It allows the derivation of Neumann-like boundary conditions
(which we call ``natural'' following Ref.~\cite{barth})
from the standard variational principle in exactly the same way
as equations of motion are derived.
(The general variation 
of the action must vanish for \emph{arbitrary} variations
of the fields in the bulk \emph{and} on the boundary.)
The second term is (a half of) the brane action which one
uses in the orbifold picture.

We present the transition from the boundary to the orbifold
picture in detail. We find that the $Y$-term disappears in
the transition. It is represented in the orbifold picture 
by brane-localized singularities of the bulk Lagrangian.

We explicitly check that the bulk-plus-brane action we obtain
for the orbifold picture is supersymmetric upon using the
same minimal set of boundary conditions as in the boundary
picture. But we find that supersymmetry of the action requires
introducing different $\ep(z)$ (sign factor) assignments for
odd fields compared to those previously assumed. Instead of
the famous $\ep(z)^2\da(z)=1/3\,\da(z)$ we find that it is
necessary to use another equally bizarre relation for a
product of distributions: $\ep(z)^{-2}\da(z)=-\da(z)$.

The construction of Ref.~\cite{bb1} is obtained from the
one presented here by explicitly using the natural boundary
conditions (including those outside the minimal set) in the
brane action and while performing the supersymmetry 
variation of the resulting bulk-plus-brane action.
We find that in the approach of Ref.~\cite{bb1} the two
alternative $\ep(z)$ assignments cannot be distinguished.

We also find that in the orbifold picture \emph{all}
local transformations have to be modified by the explicit addition
of brane-localized terms such that the resulting transformations
are non-singular on the brane. (The hint for this modification
appeared already in Refs.~\cite{abn,bb1}, where the supersymmetry
transformation of $\psi_{52}$ was modified.)
Only after this modification the orbifold picture becomes
equivalent to the boundary picture.

This paper is a companion to Ref.~\cite{my1}, where a detailed
analysis of the Mirabelli and Peskin model \cite{mp} in both the boundary
and the orbifold pictures is presented.

Our basic conventions
are the same as in Ref.~\cite{bb1}. 
We summarize them in Appendix~\ref{app-sg0}.
More details are
included in Ref.~\cite{thesis}.

\section{Supersymmetry algebra}
\label{sec-sg1}
In this section we present the (on-shell) supersymmetry algebra of 
five-dimensional gauged supergravity. The (bulk)
supergravity action and supersymmetry transformations are
as in Ref.~\cite{bb1}. But in order to show an important feature of
the local supersymmetry algebra, we need to include 3-Fermi
terms in the supersymmetry transformation of the gravitino.

The complete form of the supersymmetry transformations is
\footnote{The spinors $\Psi_{Mi}$ and $\Eta_i$ are symplectic
Majorana (see Appendix~\ref{app-sg0}). The index $i$ can be
rotated by $\Uij\in SU(2)$: $\Psi^\prime_{i}=\Uij\Psi_j$. The
(global) $SU(2)$ is the automorphism symmetry group of the
algebra when $\la\vec q=0$. The real vector 
$\vec q=(q_1, q_2, q_3)$
indicates which $U(1)$ subgroup of the $SU(2)$ has been gauged
\cite{bkvp, bb1}. One can set it to be a unit vector, $\qq=1$.}
\bea
\label{5Dsusytr}
\da_\Eta e_M^A &=& i\Etatil^i\Ga^A\Psi_{Mi} \\
\da_\Eta B_M &=& -i\frac{\sx}{2}\Etatil^i\Psi_{Mi} \\
\da_\Eta \Psi_{Mi} &=& 2 D_M(\wh\om)\Eta_i
+\frac{1}{2\sx}(\Ga_M{}^{NK}+4\da_M^N\Ga^K)\wh F_{NK}\Eta_i 
\nn\\[5pt]
\label{5Dsusytr3}
&&+\la\Qij(\Ga_M-\sx B_M)\Eta_j \; ,
\eea
where $\Qij=i(\vec q\cdot \vec\si)_i{}^j$ and
\bea
&&\wh F_{MN}=F_{MN}+i\frac{\sx}{4}\Psitil_M^i\Psi_{Ni} \\
&&\wh\om_{MAB}=\om(e)_{MAB}-\frac{i}{4}e_A^N e_B^K \left(
\Psitil_N^i\Ga_M\Psi_{Ki}+\Psitil_M^i\Ga_N\Psi_{Ki}
-\Psitil_M^i\Ga_K\Psi_{Ni} \right)
\eea
are supercovariant quantities (their 
supersymmetry variations contain no $\p_M\Eta_i$).
From these we can derive the (on-shell) supersymmetry 
algebra,\footnote{The algebra closes exactly only on the bosonic
fields $e_M^A$ and $B_M$. For the gravitino, $\Psi_{Mi}$, 
additional non-closure terms appear, proportional to its 
equation of motion. For the off-shell supersymmetry algebra
see Ref.~\cite{zu4}.}
\bea
\label{algebra}
[\da_{\rm{susy}}(\Xi), \da_{\rm{susy}}(\Eta)] =
\da_{\rm{g.c.}}(v^M)
+\da_{\rm{loc.L.}}(\om^{AB})
+\da_{U(1)}(u)
+\da_{\rm{susy}}(\Zeta_i) \; .
\eea
The commutator of two supersymmetry transformations with
parameters $\Eta_i$ and $\Xi_i$ gives a general coordinate
transformation (instead of a translation in the case of global
supersymmetry) as well as other local transformations
(in our case these are local Lorentz and $U(1)$ transformations).
But what is special to \emph{local} supersymmetry
(see, e.g., Ref.~\cite{pvn}), the commutator
also gives rise to another supersymmetry transformation!
And the 3-Fermi terms in the supersymmetry transformations
are essential to identify this feature.

A general coordinate transformation (with parameter $v^M$)
has the same form on 
$e_M^A$, $B_M$ and $\Psi_{Mi}$, since all of them carry the same
world index $M$. Explicitly, on $B_M$ it is given by
\bea
\da_v B_M =v^N\p_N B_M +B_N\p_M v^N \; .
\eea
The (local) Lorentz transformation 
(with parameter $\om_{AB}=-\om_{BA}$) is
\bea
\da_\om e_M^A = e_M^B\om_B{}^A, \qquad
\da_\om B_M = 0, \qquad
\da_\om \Psi_{Mi} = \frac{1}{4}\om_{AB}\Ga^{AB}\Psi_{Mi} \; .
\eea
And the (local) $U(1)$ transformation (with parameter $u$)
is as follows,
\bea
\label{U1tr}
\da_u e_M^A =0, \qquad
\da_u B_M =\p_M u , \qquad
\da_u \Psi_{Mi}= u \frac{\sx}{2}\la\Qij\Psi_{Mj} \; .
\eea
The parameters $v^M$, $\om_{AB}$, $u$ and $\Zeta_i$, which appear
in the commutator (\ref{algebra}), are given by
\bea
\label{saparameters}
v^M &=& 2i\Etatil^i\Ga^M\Xi_i \nn\\[5pt]
\om^{AB} &=& 2i(\Etatil^i\Ga^K\Xi_i)\wh\om_K{}^{AB}
-2i\la\Qij(\Etatil^i\Ga^{AB}\Xi_j) \nn\\[5pt]
&&-\frac{i}{\sx}\Etatil^i\Ga^{AB}{}_{NK}\Xi_i\wh F^{NK}
-\frac{4i}{\sx}\Etatil^i\Xi_i\wh F^{AB} \nn\\[5pt]
u &=& -2i(\Etatil^i\Ga^K\Xi_i)B_K-i\sx\Etatil^i\Xi_i \nn\\[5pt]
\Zeta_i &=& -i(\Etatil^j\Ga^K\Xi_j)\Psi_{Ki}  \; .
\eea

The supersymmetry algebra tells us that in order for a 
(bulk-plus-boundary) action to be supersymmetric under the
indicated supersymmetry transformations, it must also be
invariant under the local transformations arising from the 
commutator (\ref{algebra}). Namely, the general coordinate
transformation, the local Lorentz transformation and the $U(1)$
gauge transformation. This allows one to find the
boundary conditions necessary for supersymmetry of the action.

\section{Boundary breaks supersymmetry}
\label{sec-sg2}
It is well-known that in the presence of a boundary \emph{half}
of the bulk supersymmetries are necessarily broken.
The reason for this is the supersymmetry algebra, which generates
a general coordinate transformation (a translation in the case
of the global supersymmetry) in the direction normal to the boundary.

Indeed, let us consider an action on a manifold $\mc{M}$ 
with boundary $\p\mc{M}$. Its variation under the general 
coordinate transformation gives rise to a boundary term,
\bea
\label{S5gct}
\da_v S_5 
=\da_v \int_\mc{M} \mc{L}_5
=\int_\mc{M} D_M (v^M \mc{L}_5)
=\int_{\p\mc{M}} n_M v^M \mc{L}_5 \; ,
\eea
where $n^M$ is an outward pointing unit vector normal to the
boundary.
(The measures of integration, $d^5x e_5$ on $\mc{M}$
and $d^4x e_4^\text{ind}$ on $\p\mc{M}$, are implicit. ) 
This is true for
\emph{any} action, provided $\mc{L}_5$ is a scalar under
the general coordinate transformation. Thus, in the presence
of a boundary, the action is \emph{not} invariant under 
the general coordinate transformation \emph{unless}
\bea
\label{nv0}
n_M v^M =0 \qquad \text{on}\;\p\mc{M} \; .
\eea
But the supersymmetry algebra shows that a commutator of two
supersymmetry transformations generates the general coordinate
transformation with 
\bea
\label{vex}
v^M = 2i\Etatil^i\Ga^M\Xi_i \; .
\eea
And thus a restriction on $v^M$ restricts the
allowed supersymmetry transformations.

From now on we assume that the boundary is described by
\bea
\p\mc{M}:\qquad  x^5=\rm{const} \; .
\eea
The allowed general coordinate transformations,
\bea
x^M\;\rightarrow\;x^M-v^M(x) \; ,
\eea
with $n_M v^M=0$ on $\p\mc{M}$, preserve this description,
and thus our choice
does not limit the general coordinate invariance any further.
We also use a (finite) local Lorentz transformation to set 
$e_m^\hf=0$ on $\p\mc{M}$.
In this gauge, which turns out to be very convenient for our
discussion,
\bea
e_m^\hf=0, \qquad e_a^5=0, 
\qquad e_5^a\neq 0, \qquad e_\hf^m\neq 0 
\eea
and
\bea
e_m^a e_a^n=\da_m^n, \qquad 
e_a^m e_m^b=\da_a^b, \qquad
e_5^\hf e_\hf^5 =1, \qquad
e_\hf^m =-e_5^a e_a^m e_\hf^5  \; .
\eea
Note also that this is the gauge in which $e_m^a$
is a vierbein of the induced metric on the boundary (which
is not true in general) and, therefore, 
\bea
e_4^\text{ind}=e_4=\det e_m^a \; .
\eea
We also have $e_5=e_4 e_5^\hf$ and
$n_M=(0,0,0,0,n_5)$ with (see Appendix \ref{app-sg3})
\bea
n_5=-e_5^\hf \; .
\eea

After these simplifications and in the two-component
spinor notation, Eqs.~(\ref{nv0})
and (\ref{vex}) give rise to the following condition,
\bea
v^5=2i\Etatil^i\Ga^5\Xi_i
=2e_\hf^5 (\eta_2\xi_1-\eta_1\xi_2)
+h.c. =0 \qquad \text{on}\;\p\mc{M} \; .
\eea
Here $(\eta_1, \eta_2)$ and $(\xi_1, \xi_2)$
are the constituents of $\Eta_i$ and $\Xi_i$, respectively.
Clearly, its general solution is the following boundary
condition on the supersymmetry parameter,
\bea
\eta_2=\al\eta_1 \qquad \text{on}\;\p\mc{M} \; ,
\eea
where $\al$ is (for now) an arbitrary complex function of the boundary
coordinates. (We will see, however, that we can find a supersymmetric
bulk-plus-boundary action only for $\al=\rm{const}$.)
This is exactly the boundary condition used in Ref.~\cite{bb1}.
One linear combination of $\eta_1$ and $\eta_2$ gets fixed,
while the orthogonal combination describes the unbroken
supersymmetry transformation. 
We will first set $\al=0$ to simplify the discussion.\footnote{
It is sufficient to consider just the $\al=0$ case.
Any other (constant) $\al$ is obtainable by a
(global) $SU(2)$ rotation Ref.~\cite{bb1}. 
See Section \ref{sec-alpha} for details.}
Our boundary condition is then
\bea
\eta_2=0 \qquad \text{on}\;\p\mc{M} \; .
\eea
The $N=2$ supersymmetry gets broken down to
$N=1$ (described by $\eta_1$) due to the presence of the boundary.

Note that the breaking is \emph{on the boundary only}. We will find a
bulk-plus-boundary action which is invariant under the 
supersymmetry transformations
with arbitrary $\eta_1$ and $\eta_2$ in the bulk of $\mc{M}$, 
restricted only by the boundary condition on $\p\mc{M}$.
(In contrast, in the case of the \emph{global} supersymmetry, 
like in the Mirabelli and Peskin model, restricting 
\emph{constant} $\eta_1$ and $\eta_2$
on $\p\mc{M}$ is equivalent to restricting them \emph{everywhere},
and thus the $N=2$ supersymmetry is really broken down to $N=1$.)
However, in the corresponding effective four-dimensional theory
one would be able to preserve only $N=1$ supersymmetry, because
the second supersymmetry would be broken by the boundary
conditions.

\section{Boundary conditions needed for supersymmetry}
\label{sec-sg3}
Preserving the $N=1$ supersymmetry still requires some effort.
In the Mirabelli and Peskin model we found that,
in the boundary picture,
a particular boundary action is required  to preserve
the $N=1$ supersymmetry. Off-shell, no boundary condition was
necessary to establish supersymmetry of the bulk-plus-boundary
action. On-shell, however, 
some boundary conditions (which are part of the auxiliary
equations of motion) were necessary.
In our (on-shell) supergravity case, we can find
the boundary conditions that are important for supersymmetry
directly from the supersymmetry algebra!

\subsection{Boundary condition on the gravitino}
We found that the commutator of two supersymmetry transformations 
generates another supersymmetry transformation with parameter
\bea
\Zeta_i = -i(\Etatil^j\Ga^M\Xi_j)\Psi_{Mi}  \; .
\eea
In our gauge ($e_m^\hf=e_a^5=0$) and with our boundary
condition ($\eta_2=0$ on $\p\mc{M}$), we have
\bea
\Zeta_i = -e_a^m(i\eta_1\si^a\xibar_1+h.c.)\Psi_{mi} \; .
\eea
Writing $\Zeta_i$ and $\Psi_{mi}$ in terms of their 
two-component constituents, $(\zeta_1, \zeta_2)$ and
$(\psi_{m1},\psi_{m2})$, respectively, we obtain, in
particular,
\bea
\zeta_2 = -e_a^m(i\eta_1\si^a\xibar_1+h.c.)\psi_{m2} \; .
\eea
Thus, two allowed supersymmetry transformations (with $\eta_2=0$
and $\xi_2=0$) generate a \emph{forbidden} supersymmetry
transformation ($\zeta_2\neq 0$), \emph{unless} the boundary
condition
\bea
\label{psibc}
\psi_{m2}=0 \qquad \text{on}\;\p\mc{M}
\eea
is imposed. (If $\al\neq 0$, the boundary condition is 
$\psi_{m2}=\al\psi_{m1}$ on $\p\mc{M}$.)

It is important to note, however, that the supersymmetry
transformation in the commutator (\ref{algebra})
arises \emph{from the 3-Fermi terms} in the supersymmetry
variation of the gravitino. Accordingly, we expect that the boundary
condition (\ref{psibc}) is needed to prove supersymmetry of
our action to all orders in fermions, \emph{but not} just to
quadratic order in fermions.
We will show that it is indeed the case (provided an
appropriate boundary action is included).

\subsection{Boundary condition on the graviphoton}
The commutator (\ref{algebra}) also results in a $U(1)$ gauge
transformation with parameter
\bea
u = -2i(\Etatil^i\Ga^K\Xi_i)B_K-i\sx\Etatil^i\Xi_i \; .
\eea
In our gauge ($e_m^\hf=e_a^5=0$) and with our boundary
condition ($\eta_2=0$ on $\p\mc{M}$), we have
\bea
\label{uBm}
u = -2e_a^m(i\eta_1\si^a\xibar_1+h.c.)B_m \; .
\eea
This implies that there are two choices of boundary conditions
compatible with supersymmetry of the (bulk-plus-boundary) action:
\begin{enumerate}
\item
$B_m\neq 0$ on $\p\mc{M}$; the action must be $U(1)$ gauge
invariant; the boundary condition (if any) follows from
maintaining the $U(1)$ invariance.
\item
$B_m = 0$ on $\p\mc{M}$; the $U(1)$ gauge invariance is broken 
by this boundary condition and thus is not required of the action
itself; the gauge invariance must not be broken in the bulk, however,
since there the generated $u$ is non-zero.
\end{enumerate}
The first choice appears to be more attractive (gauge invariances
lead to more controlled quantum field theories), but the second
choice may also be required. In our setup this is the case
when $\la q_{12}\neq 0$.
The reason is that in this
case the boundary condition $\eta_2=0$ on $\p\mc{M}$ \emph{itself}
breaks the $U(1)$ gauge invariance! This happens because the
$U(1)$ transformation acts on the supersymmetry parameter $\Eta_i$
in the same way as it does with $\Psi_{Mi}$,
\bea
\da_u \Eta_i= u \frac{\sx}{2}\la\Qij\Eta_j \; .
\eea
For the two-component spinors this means
\bea
\da_u\eta_1 &=& i \frac{\sx}{2}\la u(q_3\eta_1-q_{12}^\ast\eta_2) \\
\da_u\eta_2 &=& i \frac{\sx}{2}\la u(-q_{12}\eta_1-q_3\eta_2) \; .
\eea
We see that the boundary condition, $\eta_2=0$ on $\p\mc{M}$,
is not invariant under this transformation, unless $\la q_{12}=0$.
(For $\al\neq 0$ the condition on $\vec q$ is accordingly modified.)

\subsection{Are there other boundary conditions?}
There are no other boundary conditions which are generated in
the way described above. The Lorentz transformation does not generate a
boundary condition since any supergravity Lagrangian is Lorentz
invariant and, therefore, no boundary term is produced when
varying the action.

This means that we should need (at most) two
boundary conditions (on the gravitino and the graviphoton) to
prove supersymmetry of the total bulk-plus-boundary action.
In particular, we should be able to do this \emph{without using any
boundary condition for the vielbein}! And we will show explicitly
that it is indeed so, provided an appropriate boundary action is
found.

We would like to emphasize that despite the limited number
of boundary conditions needed to prove supersymmetry of the total
action, a dynamical setup should include a full set
of boundary conditions to make the boundary value problem
well defined. And such a set must itself be supersymmetric (though the
use of equations of motion would be required to show the closure of
the boundary conditions under supersymmetry if one works with the
on-shell formulation of supergravity).

\section{Bulk action}
\label{sec-sg4}
In this section we introduce the bulk supergravity action
and consider its variation under the $U(1)$ gauge
and the supersymmetry transformations.

Our bulk action is the standard gauged supergravity action
in five dimensions \cite{hist}. We will omit the 4-Fermi terms and will
work to quadratic order in fermions. To this order, the action is
\bea
\label{bulkaction}
S_5 &=& \int_\mc{M} d^5\!x e_5 
\Big\{
-\frac{1}{2}R
+\frac{i}{2}\Psitil_M^i \Ga^{MNK} 
D_N \Psi_{Ki} \nn\\
&&\qquad\qquad
+ \, 6\la^2\qq
+\frac{i}{2}\Psitil_M^i \Ga^{MNK}
\left[ \frac{1}{2} \la\Qij(\Ga_N-\sx B_N) \right] \Psi_{Ki} \nn\\
&&\qquad\qquad
- i\frac{\sqrt6}{16}F_{MN}
\left(
2\Psitil^{Mi} \Psi_i^N + \Psitil_P^i \Ga^{MNPQ} \Psi_{Qi} 
\right) \nn\\
&&\qquad\qquad 
-\frac{1}{4}F_{MN}F^{MN} 
-\frac{1}{6\sqrt6}\eps^{MNPQK} F_{MN}F_{PQ}B_K 
\Big\} \; .
\eea
Note that the action with $\la\neq 0$ can be obtained from the
ungauged action (with $\la=0$) by modifying the covariant derivative
on the gravitino,
\bea
\label{mcovder}
D_M\Psi_{Ni} \quad \rightarrow \quad
\wt D_M\Psi_{Ni} =
D_M\Psi_{Ni}+\frac{1}{2}\la\Qij(\Ga_M-\sx B_M)\Psi_{Nj} \; ,
\eea
and adding the cosmological constant term $6\la^2\qq$ to
the Lagrangian. Similarly, the supersymmetry transformations
are obtained by analogous modification of the covariant
derivative on $\Eta_i$. 

The modified derivative is covariant
with respect to the $U(1)$ transformation,
\bea
\da_u (\wt D_M\Psi_{Ni})= u \frac{\sx}{2}\la\Qij
(\wt D_M \Psi_{Nj}) \; .
\eea
It is then clear that only the Chern-Simons term in the
action is not invariant under the $U(1)$ transformation 
and, therefore,
\bea
\label{S5ginv}
\da_u S_5 &=& \int_\mc{M} d^5x e_5 \Big\{
-\frac{1}{6\sqrt6}\;\eps^{MNPQK} F_{MN}F_{PQ}\da_u B_K
\Big\} \nn\\
&=& \int_\mc{M} d^5x e_5 D_K \Big[
-u \frac{1}{6\sqrt6}\;\eps^{MNPQK} F_{MN}F_{PQ} \Big] \nn\\
&=& \int_{\p\mc{M}} d^4x e_4 \Big[
-u \frac{1}{6\sqrt6}\; n_5 \eps^{5MNPQ} F_{MN}F_{PQ} \Big] \; .
\eea
In our gauge ($e_m^\hf=e_a^5=0$, $n_5=-e_5^\hf$) this becomes
\bea
\label{S5gbp}
\da_u S_5 =
\int_{\p\mc{M}} d^4x e_4 \Big[
\; u \frac{1}{6\sqrt6}\; \eps^{mnpq} F_{mn}F_{pq} \Big] \; .
\eea
Therefore, the bulk action is $U(1)$ gauge invariant
($\da_u S_5 = 0$),
provided the following (gauge invariant) boundary condition
is imposed,
\bea
F_{mn}=0 \qquad \rm{on}\;\p\mc{M} \; .
\eea
Another way to kill the boundary term is to require
$u=0$ on $\p\mc{M}$. 
The supersymmetry algebra then leads to a stronger 
(gauge non-invariant) boundary condition, 
\bea
B_m=0 \qquad \rm{on}\;\p\mc{M} \; .
\eea

Let us now consider the supersymmetry variation of the bulk action.
One can show (see the details in Ref.~\cite{thesis})
that the action varies into a boundary term (the bulk
part vanishes, since this action is known to be supersymmetric in
the absence of boundary). Explicitly,
\bea
\label{susytd}
\da_\Eta S_5 = \int_\mc{M} d^5x e_5 (D_M \wt K^M)
=\int_{\p\mc{M}} d^4x e_4 (n_M \wt K^M) \; ,
\eea
where
\bea
\wt K^M &=& -e^{MA}e^{NB}\da_\Eta\om_{NAB}
-\frac{i}{2}\Psitil_N^i\Ga^{NMK}\da_\Eta\Psi_{Ki} \nn\\
&&- F^{MN}\da_\Eta B_N
-\frac{4}{6\sx}\eps^{MNPQK}F_{PQ}B_K\da_\Eta B_N \; .
\eea
In our gauge this simplifies to
\bea
\da_\Eta S_5 
&=& \int_{\p\mc{M}} d^4x e_4 \Big\{
-e^{na}\da_\Eta\om_{na\hf}
+(\psi_{m1}\si^{mn}\da_\Eta\psi_{n2}
-\psi_{m2}\si^{mn}\da_\Eta\psi_{n1} +h.c.) \nn\\
&&\hspace{60pt} 
-e_5^\hf F^{n5}\da_\Eta B_n
+\frac{4}{6\sx}\eps^{npqk}F_{pq}B_k\da_\Eta B_n 
\Big\} \; .
\eea

\section{Boundary action for $\eta_2=0$}
\label{sec-sg5}
According to our discussion of boundary conditions necessary
for supersymmetry of the action, if we 
\begin{enumerate}
\item[1)]
work to quadratic order in fermions, and
\item[2)]
drop terms with $B_M$, 
\end{enumerate}
then there should exist
a boundary action that makes the total bulk-plus-boundary
action supersymmetric \emph{without the use of any boundary
conditions} (except the one on the supersymmetry parameter).
In this section we present such an action.

\subsection{Variational principle}
In our analysis \cite{my1}
of the Mirabelli and Peskin model \cite{mp}, we found
that the boundary action required for supersymmetry is, at the
same time, the one which makes the variational principle well
defined. Let us now turn this around. We will look for a boundary
action which improves the variational principle and then see
if it makes the total action supersymmetric.

If we consider the general variation of our action (which one would
perform to find the equations of motion), we find
\bea
\da S_5=\int_\mc{M} d^5x e_5 \Big\{
D_M K^M +(\rm{EOM})\da\Phi \Big\} \; ,
\eea
where $\Phi=\{e_M^A, B_M, \Psi_{Mi}\}$ and
\bea
K^M &=& -e^{MA}e^{NB}\da\om_{NAB}
+\frac{i}{2}\Psitil_N^i\Ga^{NMK}\da\Psi_{Ki} \nn\\
&&- F^{MN}\da B_N
-\frac{4}{6\sx}\eps^{MNPQK}F_{PQ}B_K\da B_N \; .
\eea
Note that the only difference between $K^M$ and $\wt K^M$ is in
the sign in front of the fermionic term. 
Accordingly, in our gauge ($e_m^\hf=e_a^5=0$, $n_5=-e_5^\hf$) we obtain
\bea
\da S_5 
&=& \int_{\p\mc{M}} d^4x e_4 \Big\{
-e^{na}\da\om_{na\hf}
-(\psi_{m1}\si^{mn}\da\psi_{n2}
-\psi_{m2}\si^{mn}\da\psi_{n1} +h.c.) \nn\\
&&\hspace{60pt} 
-e_5^\hf F^{n5}\da B_n
+\frac{4}{6\sx}\eps^{npqk}F_{pq}B_k\da B_n
\Big\} +(\rm{EOM}) \; .
\eea

Let us now consider the following modified action,
\bea
\label{S5pr}
S_5^\prime =S_5 + \int_{\p\mc{M}} d^4x e_4
\Big[ e^{ma}\om_{ma\hf}+ (\psi_{m1}\si^{mn}\psi_{n2}+h.c.) \Big] \; .
\eea
In our gauge $\om_{ma\hf}$ is simply related to the extrinsic
curvature (see Appendix \ref{app-sg3}),
\bea
K_{ma}=\om_{ma\hf}, \qquad K=e^{ma}\om_{ma\hf} \; .
\eea
Therefore, the first term in the boundary action is just the
standard Gibbons-Hawking term, which makes the variational problem
for the total gravity action
\bea
-\frac{1}{2}\int_\mc{M} R 
+\int_{\p\mc{M}} K
\eea
well defined. In our gauge this is especially easy to see.
The general variation of the modified action,
\bea
\label{gvS5pr}
\da S_5^\prime
&=& \int_{\p\mc{M}} d^4x e_4 \Big\{
(K_{ma}-K e_{ma})\da e^{ma}
+(2\psi_{m2}\si^{mn}\da\psi_{n1} +h.c.) \nn\\
&&\hspace{60pt} 
-e_5^\hf F^{n5}\da B_n
+\frac{4}{6\sx}\eps^{npqk}F_{pq}B_k\da B_n
\Big\} +(\rm{EOM}) \; ,
\eea
contains only the variation of the induced vierbein $e_m^a$ and
not that of its normal derivative, which is exactly 
the purpose of the
Gibbons-Hawking term. In addition, we see that the fermionic
boundary term in the modified action (\ref{S5pr}) removes 
$\da\psi_{n2}$ from the boundary piece of the general variation.
As a result, the boundary condition $\psi_{m2}=0$ on $\p\mc{M}$
(which is necessary to prove supersymmetry of the action to all 
orders in fermions) arises from requiring the general variation
to vanish under arbitrary variation $\da\psi_{m1}$ on the boundary.

The expression for the supersymmetry variation of the modified action
once again differs from the general variation (\ref{gvS5pr})
only in the fermionic piece (plus all the bulk terms are absent),
\bea
\label{svS5pr}
\da_\Eta S_5^\prime
&=& \int_{\p\mc{M}} d^4x e_4 \Big\{
(K_{ma}-K e_{ma})\da_\Eta e^{ma}
+(2\psi_{m1}\si^{mn}\da_\Eta\psi_{n2} +h.c.) \nn\\
&&\hspace{60pt} 
-e_5^\hf F^{n5}\da_\Eta B_n
+\frac{4}{6\sx}\eps^{npqk}F_{pq}B_k\da_\Eta B_n
\Big\} \; .
\eea

\subsection{Supersymmetry without boundary conditions}
Consider the following bulk-plus-boundary action,
\bea
\label{bpbaction}
S =S_5 + \int_{\p\mc{M}} d^4x e_4
\Big[ K +(\psi_{m1}\si^{mn}\psi_{n2}+h.c.) \Big]
+ \int_{\p\mc{M}} d^4x e_4 (-3\la_1) \; ,
\eea
where we added an extra tension term to the boundary action.
We now omit all $B_M$ terms and verify that the total
bulk-plus-boundary action is supersymmetric without the use of
any boundary conditions for gravitino or vierbein, but with
the restriction on the supersymmetry parameter 
$\Eta_i$,\footnote{
A similar bulk-plus-boundary action
(for a spinning string in the superconformal gauge),
supersymmetric with the use of only a boundary condition on
the supersymmetry parameter, was obtained in 1982 
in Ref.~\cite{dive1}; see their Eq.~(5.7). 
(See also Eq.~(5.17) in Ref.~\cite{dive2}.)
Analogous result for a spinning membrane, presented in a more
geometrical setting, appeared in 1989 in Ref.~\cite{lumo};
see their Eq.~(5.16).
}
\bea
\eta_2=0 \qquad \text{on}\;\p\mc{M} \; .
\eea
The supersymmetry variation gives
\bea
\da_\Eta S
= \int_{\p\mc{M}} d^4x e_4 \Big\{
-(K^{ma}-K e^{ma}+3\la_1 e^{ma})\da_\Eta e_{ma}
+(2\psi_{m1}\si^{mn}\da_\Eta\psi_{n2} +h.c.) 
\Big\},
\eea
where
\bea
\da_\Eta e_{ma} &=& -i\psi_{m1}\si^a\etabar_1 +h.c. \\[3pt]
\da_\Eta \psi_{m2} &=& 
-i K_{ma}\si^a\etabar_1+i\la q_3\si_m\etabar_1 \; .
\eea
With the help of the following identity,
\bea
K_{na}\si^{mn}\si^a =\frac{1}{2}(K_{ma}-K e_{ma})\si^a \; ,
\eea
we obtain
\bea
\da_\Eta S
= \int_{\p\mc{M}} d^4x e_4 \Big\{
3(\la_1-\la q_3)i\psi_{m1}\si^m\etabar_1 +h.c. \Big\} \; .
\eea
Therefore, the bulk-plus-boundary action (\ref{bpbaction})
is supersymmetric,
provided $\la_1=\la q_3$. And no boundary condition for the 
gravitino or vierbein is needed to prove this.

But the boundary conditions do exist for this bulk-plus-boundary
action. If we consider its general variation, we find 
\bea
\da S
&=& \int_{\p\mc{M}} d^4x e_4 \Big\{
(K_{ma}-K e_{ma}+3\la_1 e_{ma})\da e^{ma} \nn\\
&&\hspace{60pt}
+(2\psi_{m2}\si^{mn}\da\psi_{n1} +h.c.)
\Big\} +(\rm{EOM}) \; .
\eea
If we require this to vanish under arbitrary field variations,
then in addition to the bulk equations of motion we obtain the
following natural boundary conditions,
\bea
K_{ma}=\la_1 e_{ma}, \qquad
\psi_{m2}=0 \; .
\eea
If one allows the use of these boundary conditions to prove 
supersymmetry, then one can also claim that the following
bulk-plus-boundary action,
\bea
\label{correction}
S_5+ \int_{\p\mc{M}} d^4x e_4 (+\la_1) \; ,
\eea
is supersymmetric (for $\la_1=\la q_3$). (This action is 
obtained from (\ref{bpbaction}) by implementing the boundary
condition in its boundary term.) This is the approach taken
in Ref.~\cite{bb2}.\footnote{The statement in Ref.~\cite{bb2}
(see the sentence after Eq.~(2.12)) that in the
downstairs picture no boundary action is required is, in fact,
erroneous. We find here that the boundary tension term 
of Eq.~(\ref{correction}) is needed. This does not, however,
affect the rest of the analysis in Ref.~\cite{bb2}.}

Our bulk-plus-boundary action (\ref{bpbaction}) has the advantage
of giving a \emph{unique} boundary action with which the invariance
under supersymmetry is independent of the boundary conditions.
At the same time, it generates an acceptable set of
natural boundary conditions via the variational principle.

\section{Boundary action for $\eta_2=\al\eta_1$}
\label{sec-sg6}
\label{sec-alpha}
Here we show explicitly that the $\al=0$ case can be rotated into 
the $\al\neq 0$ case and that the rotated bulk-plus-boundary
action is once again supersymmetric without the use of any boundary
conditions.

\subsection{Global $SU(2)$ rotation}
We use the fact that the bulk action and the supersymmetry 
transformations are invariant under the following (global) 
rotations of the fermions ($\Psi_{Mi}$ and $\Eta_i$) 
and parameters $\vec q$,
\bea
\Psi_{Mi}^\prime =\Uij\Psi_{Mj}, \qquad
\Eta_i^\prime =\Uij\Eta_j, \qquad
Q^\prime =U Q U^\dagger \; ,
\eea
where $Q=i(\vec q\cdot \vec\si)$ and
$U$ is a constant matrix of the $SU(2)$ group.
We employ a particular rotation from this class,
\bea
\eta_1^\prime =\frac{\eta_1-\al^\ast\eta_2}{\sqrt{1+\al\al^\ast}}, 
\qquad
\eta_2^\prime =\frac{\al\eta_1+\eta_2}{\sqrt{1+\al\al^\ast}}
\eea
(similarly for $(\psi_{m1},\psi_{m2})$ and $(\psi_{51},\psi_{52})$),
together with
\bea
q_{12}^\prime = \frac
{q_{12}-\al^2 q_{12}^\ast-2\al q_3}
{1+\al\al^\ast}, \qquad
q_3^\prime = \frac
{\al q_{12}^\ast+\al^\ast q_{12}+(1-\al\al^\ast)q_3}
{1+\al\al^\ast} \; .
\eea
The inverse rotation is obtained simply by changing the sign of $\al$.
In particular,
\bea
\eta_2=\frac{-\al\eta_1^\prime+\eta_2^\prime}{\sqrt{1+\al\al^\ast}} \; .
\eea
Therefore,
\bea
\eta_2=0 \qquad \Rightarrow \qquad
\eta_2^\prime=\al\eta_1^\prime \; .
\eea

\subsection{Bulk-plus-boundary action}
Performing this rotation on the action (\ref{bpbaction})
and omitting the primes on the new fields, we obtain
\bea
\label{bpbal}
S^{(\al)} =S_5 + \int_{\p\mc{M}} d^4x e_4
\Big[ K +(\psi_{m1}\si^{mn}\psi_{n2}+h.c.) \Big]
+\int_{\p\mc{M}} d^4x e_4 \mc{L}_B^{(\al)} \; ,
\eea
where \footnote{
For a discussion of 
a boundary Lagrangian with general fermionic mass terms 
of the form $\al_{ij}\psi_i\psi_j$ see also
Ref.~\cite{gero1}.
}
\bea
\label{LBal}
\mc{L}_B^{(\al)} =
-3\la_1+(
\al_{11}\psi_{m1}\si^{mn}\psi_{n1}
+2\al_{12}\psi_{m1}\si^{mn}\psi_{n2}
+\al_{22}\psi_{m2}\si^{mn}\psi_{n2}+h.c.) \; ,
\eea
and the parameters are given by
\bea
\label{alij}
\al_{11}=\frac{-\al}{1+\al\al^\ast}, \qquad
\al_{12}=\frac{-\al\al^\ast}{1+\al\al^\ast}, \qquad
\al_{22}=\frac{\al^\ast}{1+\al\al^\ast} \; .
\eea
We claim that this bulk-plus-boundary action is supersymmetric
without the use of any boundary conditions for vierbein or gravitino
(omitting the $B_M$ terms and working to second order in fermions),
provided only that the supersymmetry parameter $\Eta_i$ is restricted
by the condition
\bea
\eta_2=\al\eta_1 \quad \rm{on}\;\p\mc{M} \; .
\eea

\subsection{Supersymmetry of the action}
The supersymmetry variation gives
\bea
\label{Salsusy}
\da_\Eta S^{(\al)}
&=& \int_{\p\mc{M}} d^4x e_4 \Big\{
-(K^{ma}-K e^{ma}+3\la_1 e^{ma})\da_\Eta e_{ma} \nn\\
&&\hspace{65pt}
+2\Big[
\big( \al_{11}\psi_{m1}+\al_{12}\psi_{m2} \big)
\si^{mn}\da_\Eta\psi_{n1} \nn\\
&&\hspace{80pt}
+\big( \al_{22}\psi_{m2}+(\al_{12}+1)\psi_{m1} \big)
\si^{mn}\da_\Eta\psi_{n2} 
+h.c. \Big]
\Big\} \; ,
\eea
where
\bea
\da e_m^a &=& -i(\psi_{m1}\si^a\etabar_1+\psi_{m2}\si^a\etabar_2) 
+h.c. \\[5pt]
\da \psi_{m1} &=& 2\wh D_m\eta_1+i K_{ma}\si^a\etabar_2
+i\la\si_m(q_3\etabar_2+q_{12}^\ast\etabar_1) \\[5pt]
\da \psi_{m2} &=& 2\wh D_m\eta_2-i K_{ma}\si^a\etabar_1
+i\la\si_m(q_3\etabar_1-q_{12}\etabar_2) \; .
\eea
Using $\eta_2=\al\eta_1$ (and assuming $\al=\rm{const}$),\footnote{
We were unable to find a bulk-plus-boundary action 
which would be supersymmetric for $\eta_2=\al\eta_1$ with
$\al$ being a \emph{function} of boundary coordinates. 
We can allow only $\al=\rm{const}$, despite the fact that
the supersymmetry algebra does not explain this limitation.} 
we can bring the variation to the 
following form,
\bea
\da_\Eta S^{(\al)}
&=& \int_{\p\mc{M}} d^4x e_4 \Big\{
4C_1\psi_{m1}\si^{mn}\wh D_n\eta_1 
+4C_2\psi_{m2}\si^{mn}\wh D_n\eta_1 \nn\\[3pt] 
&&\hspace{60pt}
+C_3(K_a^m-K e_a^m)i\psi_{m1}\si^a\eta_1 
+C_4(K_a^m-K e_a^m)i\psi_{m2}\si^a\eta_1 \nn\\[5pt]
&&\hspace{60pt}
-3C_5 i\psi_{m1}\si^m\etabar_1
-3C_6 i\psi_{m2}\si^m\etabar_1
+h.c.\Big\} \; ,
\eea
where the coefficients are
\bea
\label{Cs}
C_1 &=& \al_{11}+\al(\al_{12}+1) \nn\\
C_2 &=& \al_{12}+\al\al_{22} \nn\\
C_3 &=& \al_{11}\al^\ast-\al_{12}  \nn\\
C_4 &=& \al^\ast(\al_{12}+1)-\al_{22} \nn\\
C_5 &=& -\la_1+\la A_1 \nn\\
C_6 &=& -\la_1\al^\ast+\la A_2
\eea
with
\bea
A_1 &=& \al_{11}(q_3\al^\ast+q_{12}^\ast)
+(\al_{12}+1)(q_3-q_{12}\al^\ast) \\[5pt]
A_2 &=& \al_{12}(q_3\al^\ast+q_{12}^\ast)
+\al_{22}(q_3-q_{12}\al^\ast)  \; .
\eea
Our bulk-plus-boundary action is supersymmetric without
the use of boundary conditions for the fields if all 
the coefficients $C_i$ vanish. And indeed, using our expressions
(\ref{alij}) for the parameters $\al_{ij}$, we find
\bea
C_1=C_2=C_3=C_4=0  \; .
\eea
We also find $A_2=\al^\ast A_1$ as well as 
\bea
A_1=-\frac{\al q_{12}^\ast+\al^\ast q_{12}+(\al\al^\ast-1)q_3}
{1+\al\al^\ast} \; .
\eea
Therefore, the remaining conditions, $C_5=0$ and $C_6=0$, are
satisfied provided we choose $\la_1$ in the boundary Lagrangian
(\ref{LBal}) to be
\bea
\label{laal}
\la_1=-\frac{\al q_{12}^\ast+\al^\ast q_{12}+(\al\al^\ast-1)q_3}
{1+\al\al^\ast}\la \; .
\eea
This is exactly the relation found in Ref.~\cite{bb1}.
(Note also that it is just the rotated version of $\la_1=\la q_3$
for $\al=0$.) But, unlike Ref.~\cite{bb1} or \cite{bb2}, 
we did not have to use 
boundary conditions to prove supersymmetry of the total action.

\subsection{Boundary conditions}
The general variation of the bulk-plus-boundary action (\ref{bpbal})
gives
\bea
\da S
&=& \int_{\p\mc{M}} d^4x e_4 \Big\{
(K_{ma}-K e_{ma}+3\la_1 e_{ma})\da e^{ma} \nn\\
&&\hspace{60pt}
+2\Big[
\big( \al_{11}\psi_{m1}+(\al_{12}+1)\psi_{m2} \big)
\si^{mn}\da\psi_{n1} \nn\\
&&\hspace{80pt}
+(\al_{12}\psi_{m1}+\al_{22}\psi_{m2})
\si^{mn}\da\psi_{n2} 
+h.c. \Big]
\Big\} +(\rm{EOM}) \; .
\eea
One could worry that having both $\da\psi_{m1}$ and 
$\da\psi_{m2}$ in the boundary piece of the general variation
could interfere with the application of the variational principle.
And indeed, if we require the variation to vanish for arbitrary
$\da e^{ma}$, $\da\psi_{m1}$ and $\da\psi_{m2}$ on the boundary,
we would obtain \emph{two} fermionic boundary conditions,
\bea
\psi_{m2}=-\frac{\al_{11}}{\al_{12}+1}\psi_{m1}, \qquad
\psi_{m2}=-\frac{\al_{12}}{\al_{22}}\psi_{m1} \; .
\eea
For general $\al_{ij}$ this would overdetermine the boundary
value problem. However, for our special choice (\ref{alij}),
the two boundary conditions reduce to one!
This saves the variational principle.

Our bulk-plus-boundary action leads to the following
natural boundary conditions,
\bea
K_{ma}=\la_1 e_{ma}, \qquad \psi_{m2}=\al\psi_{m1} \; .
\eea
These, once again, coincide with the boundary conditions used
in Refs.~\cite{bb1} and \cite{bb2}. 
And if we plug these boundary conditions
in our boundary Lagrangian (\ref{LBal}), we find
\bea
\label{shortLB}
\mc{L}_B^{(\al)}=-3\la_1-(\al\psi_{m1}\si^{mn}\psi_{n1}+h.c.) \; ,
\eea
which is (a half of) the brane Lagrangian in Ref.~\cite{bb1}.

\subsection{$U(1)$ gauge invariance}
We now investigate when the boundary action 
and/or boundary conditions are gauge invariant. 
The $U(1)$ gauge transformation on the fermions is
\bea
\da_u\eta_1 &=& i w(q_3\eta_1-q_{12}^\ast\eta_2) \nn\\[5pt]
\da_u\eta_2 &=& i w(-q_3\eta_2-q_{12}\eta_1) \; ,
\eea
where $w\equiv i\frac{\sx}{2}\la u \in\mathbb{R}$; the 
transformation on the $\psi_{m1,2}$ is similar.

The boundary condition $\eta_2=\al\eta_1$ is gauge invariant if
\bea
\da_u(\eta_2-\al\eta_1)=-i w\big[
(q_{12}+\al q_3)-\al(\al q_{12}^\ast-q_3) \big] =0 \; .
\eea
This leads to the following quadratic equation for $\al$,
\bea
q_{12}-\al^2 q_{12}^\ast+2\al q_3 =0 \; .
\eea
It is equivalent to two linear equations,
\bea
\label{ginval}
\al q_{12}^\ast-q_3=\pm 1, \qquad
q_{12}+\al q_3 =\pm\al \; ,
\eea
where the signs $\pm$ correspond to the two solutions of the
quadratic equation \cite{bb1}. (We assumed here the normalization
condition $\qq=1$ of Ref.~\cite{bb1}.)

The fermionic part of the boundary Lagrangian 
(\emph{including} the $\psi_{m1}\si^{mn}\psi_{n2}$ term) is
\bea
\mc{L}_{BF}=\frac{1}{1+\al\al^\ast}
\Big[ 
-\al\psi_1\psi_1
+(1-\al\al^\ast)\psi_1\psi_2
+\al^\ast\psi_2\psi_2
\Big] \; ,
\eea
where we used a shorthand notation 
$\psi_i\psi_j\equiv\psi_{mi}\si^{mn}\psi_{nj}$ and dropped
the ``$+h.c.$''.
(Note that $\mc{L}_{BF}$ \emph{vanishes} when the
boundary condition $\psi_{m2}=\al\psi_{m1}$ is used!)
Its variation under the gauge transformation gives
\bea
\label{gvarLB}
\da_u \mc{L}_{BF}=\frac{i w}{1+\al\al^\ast}
\Big[ 
W\psi_1\psi_1
+2(\al q_{12}^\ast -\al^\ast q_{12})\psi_1\psi_2
+W^\ast\psi_2\psi_2
\Big] \; ,
\eea
where
\bea
W=\al(\al^\ast q_{12}-q_3)-(\al q_3+ q_{12}) \; .
\eea
We see that the variation vanishes when $\al$ and $\vec q$
are related as in Eq.~(\ref{ginval}). We conclude, therefore,
that the (full) boundary Lagrangian and the boundary condition 
$\eta_2=\al\eta_1$
are \emph{both} gauge invariant when the equation (\ref{ginval})
is satisfied! 

For future reference, we note that if we do not include the 
$\psi_{m1}\si^{mn}\psi_{n2}$ term,
that is if we consider the variation of
\bea
\mc{L}_{BF}^{(\al)}=\frac{1}{1+\al\al^\ast}
\Big[ 
-\al\psi_1\psi_1
-2\al\al^\ast\psi_1\psi_2
+\al^\ast\psi_2\psi_2
\Big]
\eea
under the gauge transformation, we find
\bea
\label{gvarLBal}
\da_u \mc{L}_{BF}^{(\al)}=\frac{i w}{1+\al\al^\ast}
\Big[ 
\wt W\psi_1\psi_1
+2(\al q_{12}^\ast -\al^\ast q_{12})\psi_1\psi_2
+\wt W^\ast\psi_2\psi_2
\Big] \; ,
\eea
where 
\bea
\wt W=2 \al(\al^\ast q_{12}-q_3) \; .
\eea
Therefore, when the boundary condition $\eta_2=\al\eta_1$ is
gauge invariant, i.e. the equation (\ref{ginval}) is satisfied,
the $\al$-dependent part of the boundary Lagrangian, 
$\mc{L}_B^{(\al)}$,
is \emph{not} gauge invariant! (Unless $\al=0$, in which
case $\mc{L}_{BF}^{(\al)}=0$; or $\la=0$, in which case the
$U(1)$ does not act on the fermions.)

\section{Fate of $B_M$ terms}
\label{sec-sg7}
In this section we will keep all $B_M$ and $F_{MN}$ terms 
(as well as $e_5^a$ and $e_\hf^m$), 
and show that, with an appropriate addition
to the boundary action, the bulk-plus-boundary action is 
supersymmetric provided we use the (gauge non-invariant)
boundary condition $B_m=0$ on $\p\mc{M}$.

\subsection{Old action}
Consider the modified bulk action (\ref{S5pr}),
\bea
S_5^\prime =S_5 + \int_{\p\mc{M}} d^4x e_4
\Big[ K + (\psi_{m1}\si^{mn}\psi_{n2}+h.c.) \Big] \; .
\eea
Its general variation is (\ref{gvS5pr}),
\bea
\da S_5^\prime
&=& \int_{\p\mc{M}} d^4x e_4 \Big\{
-(K_a^m-K e_a^m)\da e_m^a
+(2\psi_{m2}\si^{mn}\da\psi_{n1} +h.c.) \nn\\
&&\hspace{60pt} 
-e_5^\hf F^{n5}\da B_n
+\frac{4}{6\sx}\eps^{npqk}F_{pq}B_k\da B_n
\Big\} +(\rm{EOM}) \; ,
\eea
and its supersymmetry variation is given by (\ref{svS5pr}),
\bea
\da_\Eta S_5^\prime
&=& \int_{\p\mc{M}} d^4x e_4 \Big\{
-(K_a^m-K e_a^m)\da_\Eta e_m^a
+(2\psi_{m1}\si^{mn}\da_\Eta\psi_{n2} +h.c.) \nn\\
&&\hspace{60pt} 
-e_5^\hf F^{n5}\da_\Eta B_n
+\frac{4}{6\sx}\eps^{npqk}F_{pq}B_k\da_\Eta B_n
\Big\} \; .
\eea
This is true for any $\Eta_i$.

\subsection{Supersymmetry with $B_M$ terms}
Let us first discuss the case $\al=0$, so that the boundary condition
on the supersymmetry parameter $\Eta_i$ is
\bea
\eta_2=0 \qquad \rm{on}\;\p\mc{M} \; .
\eea
The supersymmetry transformations then are 
\bea
\da_\Eta e_m^a &=& -i\psi_{m1}\si^a\etabar_1+h.c. \\
\da_\Eta B_m &=& i\frac{\sx}{2}\psi_{m2}\eta_1+h.c.\\[3pt]
\da_\Eta \psi_{m2} &=& 
-i K_{ma}\etabar_1+i\la q_3\si_m\etabar_1 \nn\\[3pt]
&&+i\sx\la q_{12}\eta_1 B_m -\frac{1}{2\sx}
(i\eps_m{}^{nkl}\si_l+4\da_m^n\si^k)\etabar_1 F_{nk} \; .
\eea
Therefore,
\bea
\da_\Eta S_5^\prime
&=& \int_{\p\mc{M}} d^4x e_4 \Big\{
-3i\la q_3 (\psi_{m1}\si^m\etabar_1) \nn\\
&&+2i\sx\la q_{12} (\psi_{m1}\si^{mn}\eta_1) B_n
-\frac{3i}{2\sx}(\psi_{m1}\si_n\etabar_1)\eps^{mpqn}F_{pq} \nn\\
&&-i\frac{\sx}{2}(\psi_{n2}\eta_1)e_5^\hf F^{n5}
+\frac{i}{3}(\psi_{n2}\eta_1)\eps^{npqk}F_{pq}B_k +h.c.
\Big\} \; .
\eea
The term with $\la q_3$ can be compensated by a boundary tension term
(see Eq.~(\ref{bpbaction})). Most of the other terms can be killed
by application of the boundary condition for $B_m$,
\bea
F_{mn}=0 \qquad \rm{on}\;\p\mc{M}
\eea
if $\la q_{12}= 0$ and the gauge invariance is preserved on the
boundary; or 
\bea
B_m=0 \qquad \rm{on}\;\p\mc{M}
\eea
if $\la q_{12}\neq 0$ and the gauge invariance on the boundary is
broken by the boundary condition on the supersymmetry parameter.
However, we are still left with
\bea
\label{remnant}
\da_\Eta S_5^\prime
= \int_{\p\mc{M}} d^4x e_4 \Big\{
-i\frac{\sx}{2}(\psi_{m2}\eta_1)e_5^\hf F^{n5}
\Big\} \; .
\eea

This remaining term can, in principle, be canceled by the
boundary condition on the gravitino,
\bea
\psi_{m2}=0 \qquad \rm{on}\;\p\mc{M} \; .
\eea
However, our analysis of the supersymmetry algebra indicates that
we \emph{should not need} this boundary condition for 
supersymmetry to \emph{quadratic order} in fermions.
Therefore, there should exist a boundary action which lets us
avoid using this boundary condition. 

Another reason in favor of modifying our action
is that, at the moment, the $B_m$ boundary
condition we need ($B_m=0$ or $F_{mn}=0$
on $\p\mc{M}$) is not the same as the natural boundary 
condition arising from the bulk-plus-boundary action,
\bea
e_5^\hf F^{n5}
-\frac{4}{6\sx}\eps^{npqk}F_{pq}B_k =0 \qquad \rm{on}\;\p\mc{M} \; .
\eea

\subsection{New action}
In order to 
\begin{enumerate}
\item[1)]
avoid using the $\psi_{m2}=0$ boundary condition
in proving supersymmetry (to quadratic order in fermions), and 
\item[2)]
have the boundary condition for $B_m$ ($B_m=0$ or $F_{mn}=0$
on $\p\mc{M}$) appear as a natural boundary condition,
\end{enumerate}
we need to add an appropriate boundary action. Such
a boundary action exists, but it turns out that it itself
breaks gauge invariance! That is, for agreement with the 
supersymmetry algebra we have to break the gauge invariance
(only on the boundary) by hand. However, we will find that
the same boundary action is also needed for the correct transition to
the orbifold picture.

The required boundary action is easy to find. Indeed,
consider the following bulk-plus-boundary action,
\bea
\label{S5prpr}
S_5^{\prime\prime}= S_5^\prime +\int_{\p\mc{M}} d^4x e_4 \Big\{
e_5^\hf F^{n5} B_n \Big\} \; .
\eea
Its general variation is,
\bea
\da S_5^{\prime\prime}
&=& \int_{\p\mc{M}} d^4x e_4 \Big\{
-[K_a^m-K e_a^m-(B_n C^n) e_a^m]\da e_m^a
+(2\psi_{m2}\si^{mn}\da\psi_{n1} +h.c.) \nn\\
&&\hspace{60pt} +B_m \da C^m
+\frac{4}{6\sx}\eps^{npqk}F_{pq}B_k\da B_n
\Big\} +(\rm{EOM}) \; ,
\eea
where we defined
\bea
C^m\equiv e_5^\hf F^{m5} &=& e_5^\hf \big[
g^{mn}g^{k5}F_{nk}+(g^{mn}g^{55}-g^{m5}g^{n5})F_{n5} \big] 
\nn\\[5pt]
&=& \ga^{mn}F_{n\hf} \equiv e_a^m e^{na}
(e_\hf^5 F_{n5}+e_\hf^k F_{nk}) \; .
\eea
The natural boundary conditions corresponding to \emph{independent}
variations of $B_m$ and $C^m$ 
(they are independent since $C^m$ involves $\p_5 B_n$ whose
value on $\p\mc{M}$ is independent of the value of $B_n$)
coincide on the following boundary
condition,
\bea
B_m =0 \qquad \rm{on}\;\p\mc{M} \; .
\eea
This is exactly the (gauge non-invariant) boundary condition
dictated by the supersymmetry algebra. (Note that the $B_m$
and $C^m$ fields are analogous to, respectively, 
the $\Phi$ and $D=X_3-\p_5\Phi$ fields
of the Mirabelli and Peskin model \cite{mp}. There, in order to derive
the boundary condition for $\Phi$, we need a boundary term
$\Phi D$ \cite{my1}.)

The result of the supersymmetry variation, 
compared to Eq.~(\ref{remnant}), is now
\bea
\da_\Eta S_5^{\prime\prime}
= \int_{\p\mc{M}} d^4x e_4 \Big\{
B_n\da_\Eta C^n +(B_n C^n)e_a^m \da_\Eta e_m^a
\Big\} \; .
\eea
And, therefore, the bulk-plus-boundary action (\ref{S5prpr})
is supersymmetric upon using only the $B_m=0$ boundary
condition, but not the $\psi_{m2}=0$ one.

\subsection{Extension to the $\al\neq 0$ case}
The generalization to the $\al\neq 0$ case is straightforward.
Since $B_M$ is not rotated under the $SU(2)$, we do not get
new boundary terms for the $\al\neq 0$ case. The boundary
condition remains $B_m=0$ on $\p\mc{M}$. The only terms we
should consider, therefore, are those containing $F_{m5}$.

The variation (\ref{Salsusy}) can also be written as follows,
\bea
\da_\Eta S^{(\al)}
&=& \int_{\p\mc{M}} d^4x e_4 \Big\{
-(K^{ma}-K e^{ma}+3\la_1 e^{ma})\da_\Eta e_{ma} \nn\\
&&\hspace{65pt}
+2\Big[
\frac{\psi_{m1}+\al^\ast\psi_{m2}}{1+\al\al^\ast}
\si^{mn}(\da_\Eta\psi_{n2}-\al\da_\Eta\psi_{n1})
+h.c. \Big]
\Big\} \; .
\eea
The terms with $F_{m5}$ appear only in the variation of
the gravitino,
\bea
\da_\Eta \psi_{m1} &=& \frac{-4i}{2\sx}
(\si_m{}^n+\da_m^n)\eta_1
(e_\hf^5 F_{n5}) +\dots \\
\da_\Eta \psi_{m2} &=& \frac{-4i}{2\sx}
(\si_m{}^n+\da_m^n)\eta_2
(e_\hf^5 F_{n5}) +\dots \; ,
\eea
where dots represent the terms already considered.
But since $\eta_2=\al\eta_1$ on $\p\mc{M}$, the $F_{m5}$
terms cancel in the combination 
$\da_\Eta\psi_{n2}-\al\da_\Eta\psi_{n1}$ !
(Note that in the orbifold picture the cancellation of
the $F_{m5}$ terms is less straightforward, as we will
see.)
This completes our explicit check of the fact that 
the rotated action,
\bea
S^{(\al)}+\int_{\p\mc{M}} d^4x e_4 \Big\{
e_5^\hf F^{n5} B_n \Big\} \; ,
\eea
is supersymmetric (using only the $\eta_2=\al\eta_1$ and 
$B_m=0$ boundary conditions).

\section{From boundary to orbifold picture}
\label{sec-sg8}
\label{sec-trans}
In this section we will show that that the generalized
Gibbons-Hawking term \cite{gibh} 
(which we call ``$Y$-term'' to honor
the work of York \cite{york1, york2}) 
matches onto the brane-localized singularities
of the bulk Lagrangian in the orbifold picture. This explains
why the $Y$-term appears only in the bulk-plus-boundary action
(in the boundary picture),
but not in the bulk-plus-brane action
(in the orbifold picture).

\subsection{Summary of the boundary picture discussion}
Our total bulk-plus-boundary action is
\bea
S=\int_\mc{M} d^5x e_5 \mc{L}_5
+\int_{\p\mc{M}} d^4x e_4 Y
+\int_{\p\mc{M}} d^4x e_4 \mc{L}_B^{(\al)} \; ,
\eea
where the $\al$-independent boundary term $Y$ is 
\bea
\label{Yterm}
Y=K+(\psi_{m1}\si^{mn}\psi_{n2}+h.c.)+e_5^\hf F^{n5} B_n 
\eea
and $\mc{L}_B^{(\al)}$ is given by Eq.~(\ref{LBal})
together with Eqs.~(\ref{alij}) and (\ref{laal}).
We showed that
the action is supersymmetric for $\eta_2=\al\eta_1$ on $\p\mc{M}$
provided we use just one more boundary condition: 
$B_m=0$ on $\p\mc{M}$.

The $Y$-term is a generalization of the Gibbons-Hawking
boundary term for our bulk action. It allows us to derive
(natural) boundary conditions by requiring that the
general variation of the action vanish for arbitrary field
variations in the bulk and on the boundary. The boundary
conditions thus obtained are
\bea
\label{bybc}
B_m=0, \qquad K_{ma}=\la_1 e_{ma}, \qquad
\psi_{m2}=\al\psi_{m1} \quad \rm{on}\;\p\mc{M} \; .
\eea
They are consistent with supersymmetry, as was shown in 
Ref.~\cite{bb1}.
Their supersymmetry variations also produce other (secondary)
boundary conditions. In on-shell formulation, one can at most
expect that the full system of boundary conditions closes
under supersymmetry only \emph{up to equations of motion} 
\cite{my1}. (See also Ref.~\cite{nieu}.)

\subsection{Lifting to the orbifold}
Let us now lift our results to the $\mathbb{R}/\mathbb{Z}_2$ 
orbifold. The  orbifold
can be constructed from two copies of our space-time manifold
with boundary,
\bea
\mc{M}_-=\mathbb{R}^{1,3}\times(-\infty, 0] 
\quad \rm{and} \quad
\mc{M}_+=\mathbb{R}^{1,3}\times[0, +\infty) \; .
\eea
(We use the notation valid for the flat case \cite{my1}, but keep
in mind that we are actually on a general curved manifold.)
Since the boundaries of $\mc{M}_+$ and $\mc{M}_-$ coincide, we denote
$\Si=\p\mc{M}_+=\p\mc{M}_-$ and call this hypersurface a ``brane''.

We use a $\mathbb{Z}_2$ symmetry of the bulk action to
impose the following parity assignments \cite{bb1}
(``even/odd'' means $\phi(-z)=\pm\phi(+z)$),
\bea
\bma
\rm{even}: & e_m^a & e_5^\hf & B_5 & \psi_{m1} & \psi_{52} & \eta_1 
& q_{12} & \la \\[5pt]
\rm{odd}: & e_m^\hf & e_5^a & B_m & \psi_{m2} & \psi_{51} & \eta_2
& q_3  \; . & {}
\ema
\eea
It follows that $K=e^{ma}\om_{ma\hf}$ is odd and $F^{m5}$ is even.
Therefore, the $Y$-term is odd.

Since we used $n_5=-e_5^\hf$, our bulk-plus-boundary action is
appropriate for $\mc{M}_+$. For $\mc{M}_-$, we should use 
$n_5=+e_5^\hf$ (see Appendix \ref{app-sg2}) and accordingly change 
the sign of the $Y$-term. The boundary conditions, which our actions
for $\mc{M}_+$ and $\mc{M}_-$ should reproduce, are
\bea
&B_m^{(+)}=0, \qquad K_{ma}^{(+)}=+\la_1^{(+)} e_{ma}, \qquad
\psi_{m2}^{(+)}=+\al^{(+)}\psi_{m1}& \\
&B_m^{(-)}=0, \qquad K_{ma}^{(-)}=-\la_1^{(+)} e_{ma}, \qquad
\psi_{m2}^{(-)}=-\al^{(+)}\psi_{m1}&.
\eea
(The superscripts $(\pm)$ mean 
``evaluated on the $\mc{M}_\pm$ side of the brane $\Si$''.)
This is equivalent to using Eq.~(\ref{bybc}) on both sides of $\Si$
if we set $\al^{(-)}=-\al^{(+)}$ and $\la_1^{(-)}=-\la_1^{(+)}$.
From the expression for $\la_1$, Eq.~(\ref{laal}), we see that
\bea
\al^{(-)}=-\al^{(+)}, \quad q_3^{(-)}=-q_3^{(+)}
\quad \Rightarrow \quad
\la_1^{(-)}=-\la_1^{(+)} \; .
\eea
We, therefore, find that our boundary Lagrangian $\mc{L}_B^{(\al)}$ is
\emph{odd}. (In the Mirabelli and Peskin model, the boundary
Lagrangian $\mc{L}_4$ includes only even bulk fields and is
itself even. Our $\mc{L}_B^{(\al)}$, however, includes also
an odd bulk field, $\psi_{m2}$, and odd parameters, $\al$ and $q_3$.)

The correct actions for the both sides then are
\bea
S_\pm =\int_{\mc{M}_\pm} d^5x e_5 \mc{L}_5
\pm \int_{\p\mc{M}_\pm} d^4x e_4 Y^{(\pm)}
\pm \int_{\p\mc{M}_\pm} d^4x e_4 \mc{L}_B^{(\al)}{}^{(\pm)} \; .
\eea
The orbifold action is the sum of these two,
\bea
\label{pmsum}
S=\int_{\mc{M}_+\cup\mc{M}_-} d^5x e_5 \mc{L}_5
+\int_\Si d^4x e_4 [2Y^{(+)}]
+\int_\Si d^4x e_4 [2\mc{L}_B^{(\al)}{}^{(+)}] \; .
\eea
By analogy with our analysis of
the Mirabelli and Peskin model \cite{my1}, we expect that
the $Y$-term matches onto the brane-localized terms produced by the
bulk Lagrangian $\mc{L}_5$.
We will show now that the match is (almost) perfect.

\subsection{Boundary $Y$-term vs. Orbifold singular terms}
For each odd field, we can write
\bea
\Phi(x,z)=\ep(z)\Phi^{(+)}(x,|z|), \qquad
\p_5\Phi=(\p_5\Phi)^{(+)}+2\Phi^{(+)}\da(z) \; ,
\eea
where $\ep(z)=\pm 1$ on $\mc{M}_\pm$. In particular,
\bea
F_{m5}=F_{m5}^{(+)}-2B_m^{(+)}\da(z) \; .
\eea
This allows us to separate the $\Si$-localized terms in $\mc{L}_5$
explicitly.

The relevant part of the bulk Lagrangian is
\bea
\mc{L}_5 &=& -\frac{1}{2}R
+\frac{i}{2}\Psitil_M^i\Ga^{M5K}\p_5\Psi_{Ki} \nn\\
&& -\frac{1}{4}F_{MN}F^{MN}
-\frac{1}{6\sx}\eps^{MNPQK}F_{MN}F_{PQ}B_K
+\dots
\eea
The analysis simplifies a lot in our gauge 
($e_m^\hf=e_a^5=0$).\footnote{Note that we keep 
$e_5^a\neq 0$, $e_\hf^m\neq 0$. See Appendix~\ref{app-sg3}.}
In particular, a HUGE advantage of this gauge is that
\bea
\fbox{
\text{there are no $\da(z)$-terms in $\om_{MAB}$ !}
} \nn
\eea
Therefore,
\bea
R &=& -2e^{5\hf} e^{ma}\p_5\om_{ma\hf}+\dots \nn\\[5pt]
&=& -4e^{5\hf}e^{ma}\om_{ma\hf}^{(+)}\da(z)+\dots \nn\\[5pt]
&=& -4e_\hf^5 K^{(+)}\da(z)+\dots \; ,
\eea
where the dots denote non-singular terms. Next,
\bea
\frac{i}{2}\Psitil_M^i\Ga^{M5K}\p_5\Psi_{Ki}
&=& e_\hf^5[\psi_{m1}\si^{mn}\p_5\psi_{n2}+h.c.]+\dots \nn\\
&=& e_\hf^5[2\psi_{m1}\si^{mn}\psi_{n2}^{(+)}\da(z)+h.c.]+\dots
\eea
The Chern-Simons term is straightforward to consider,
\bea
\eps^{MNPQK}F_{MN}F_{PQ}B_K
&=& -4e_\hf^5\eps^{mpqk}F_{m5}F_{pq}B_k +\dots \nn\\
&=& 8e_\hf^5\eps^{mpqk}F_{pq}B_k B_m^{(+)}\da(z)+\dots \nn\\
&=& 8e_\hf^5\eps^{mpqk}
F_{pq}^{(+)}B_k^{(+)} B_m^{(+)}\ep(z)^2\da(z)+\dots \nn\\
&=& 0+\dots \; ,
\eea
whereas the $B_M$ kinetic term is the trickiest to analyze. We have
\bea
-\frac{1}{4}F_{MN}F^{MN}=-\frac{1}{4}F_{mn}F^{mn}
-\frac{1}{2}F_{m5}F^{m5} \; .
\eea
\emph{Both} terms contain $\da(z)$-terms. 
Indeed, let us introduce
\bea
\al^{mnk}=g^{mn}g^{k5}, \qquad 
\beta^{mn}=g^{mn}g^{55}-g^{m5}g^{n5} \; .
\eea
Then
\bea
F^{m5}=\al^{mnk}F_{nk}+\beta^{mn}F_{n5}, \qquad
F^{mn}=g^{mk}g^{nl}F_{kl}+(\al^{mkn}-\al^{nkm})F_{k5} \; ,
\eea
and we find
\bea
-\frac{1}{4}F_{mn}F^{mn}
&=& (\al^{mkn}F_{mn}B_k)^{(+)}\ep(z)^2\da(z) 
+\dots \\
-\frac{1}{2}F_{m5}F^{m5}
&=& 2\beta^{mn}F_{m5}^{(+)}B_n^{(+)}\da(z)
+(\al^{mkn}F_{mn}B_k)^{(+)}\ep(z)^2\da(z) \nn\\
&& -2\beta^{mn}B_m^{(+)}B_n^{(+)}\da(z)^2 +\dots
\eea
Finally, combining the pieces, we obtain the following expression
for the singular part of the bulk Lagrangian
(the $e_5^\hf$ is taken from $e_5=e_4 e_5^\hf$),
\bea
e_5^\hf \mc{L}_5 &=& 2K^{(+)}\da(z)
+[2\psi_{m1}\si^{mn}\psi_{n2}^{(+)}+h.c.]\da(z) \nn\\[5pt]
&& +e_5^\hf\Big\{\;
2\beta^{mn}F_{m5}^{(+)}B_n^{(+)}\da(z)
+2(\al^{mkn}F_{mn}B_k)^{(+)}\ep(z)^2\da(z) \nn\\[3pt]
&&\hspace{127pt}
-2\beta^{mn}B_m^{(+)}B_n^{(+)}\da(z)^2\Big\} +\dots
\eea
This is to be compared with
\bea
2 Y^{(+)}\da(z)
&=& 2K^{(+)}\da(z)
+2[\psi_{m1}\si^{mn}\psi_{n2}+h.c.]^{(+)}\da(z) \nn\\[5pt]
&& +2e_5^\hf \Big[\beta^{mn}F_{m5}B_n 
+\al^{mkn}F_{mn}B_k \Big]^{(+)}\da(z) \; .
\eea

\subsection{Auxiliary boundary condition}
\label{secaux}
We see that we definitely do not match the $\da(z)^2$ terms.
To do so, one would have to put $\da(0)$ terms on the boundary
which we consider unnatural. Instead, we refer to the discussion
of the Mirabelli and Peskin model \cite{my1}, where it was found that 
the $\da(z)^2$ terms are taken care of by the auxiliary fields 
upon going on-shell. 

How could this help if there are no auxiliary fields, but
$\da(z)^2$ terms are present?
The point is that ``going on-shell'' in the boundary picture
means not only eliminating the auxiliary fields, but also using
some boundary conditions which are a part of the auxiliary equations
of motion \cite{my1}.

We conjecture that $B_m=0$ on $\p\mc{M}$ is exactly such
an ``auxiliary boundary condition''. (This is so if, 
in the $Y$-term for the \emph{off-shell} supergravity action,
the $B_m$ appears multiplied by an auxiliary field.)
Using this boundary condition takes care of the discrepancy
in the $\da(z)^2$ terms.

\subsection{Different $\ep(z)$ for different fields}
\label{secDifEp}
The other mismatch is in the term with $\ep(z)^2$,
\bea
\label{epmis}
2(\al^{mkn}F_{mn}B_k)^{(+)}\ep(z)^2\da(z) \; \in \; \mc{L}_5 \; .
\eea
Setting $\ep(z)^2=1$ would eliminate the discrepancy, but
we are not allowed to do so, since 
\footnote{This relation
was first noticed by Conrad in Ref.~\cite{conrad}.
See also Ref.~\cite{bds}. The key to its understanding is the
fact that the ``sign function'' $\ep(z)$ must be treated as
a \emph{distribution}, just like the delta function $\da(z)$.
One way to define it is via a limit of a sequence of regular
(smooth) functions: $\ep(z)=\lim\ep_n(z)$. Accordingly,
$\da(z)=\lim\da_n(z)$. The product of distributions is ill-defined
\emph{unless} we relate the two sequences. We require
$\ep_n^\prime(z)=2\da_n(z)$. 
Then $\lim \int dz \ep_n^2(z)\da_n(z) f(z)=(1/3)\lim \int dz\da_n(z) f(z)$
for any (smooth) test function $f(z)$. This gives precise
meaning to the distributional equality $\ep(z)^2\da(z)=(1/3)\da(z)$.}
\bea
\label{ep2}
\ep(z)^2\da(z)=\frac{1}{3}\da(z) \; .
\eea

One could find various ``excuses'' for neglecting this term.
One could use the $B_m=0$ boundary condition to kill it.
But this also kills the $\beta^{mn}F_{m5}B_n$ term which matches 
perfectly. Or one could argue that it is ``of higher order in the brane
coupling''. Indeed, this term is special in the sense
that it is a product of three odd fields
($g^{m5}$, $F_{mn}$ and $B_m$) evaluated on $\p\mc{M}_+$.
If the brane action is such that
these fields acquire non-zero boundary conditions, this
term becomes proportional to $g^3$, with $g$ being a coupling constant
in front of the boundary action. (This is exactly the type of
expansion used by Horava and Witten in Ref.~\cite{hw}.
The role of $g$ is played there by $\kappa^{2/3}$.)

But there is, actually, another way to eliminate the mismatch. 
And it can be motivated as follows.
Note that the orbifold construction may correspond to a discontinuous
limit of some smooth supergravity realization
(when the brane sources are smoothed out into the bulk).
The $\ep(z)$ would then correspond to a smooth warp-factor.
But then, why would all the odd fields have the same warp-factor?

Let us, therefore, introduce different $\ep(z)$ for different
odd fields! We numerate them as follows,
\bea
&&\eta_2=\ep_1\eta_2^{(+)}, \quad
\psi_{m2}=\ep_2\psi_{m2}^{(+)}, \quad
K_{ma}=\ep_3 K_{ma}^{(+)}, \quad
q_3=\ep_4 q_3^{(+)} \\[5pt]
&&B_m=\ep_5 B_m^{(+)}, \quad
e_{5a}=\ep_6 e_{5a}^{(+)} \; .
\eea
They have to satisfy $\ep_i(z)=\pm 1$ on $\mc{M}_\pm$,
but if we have one such $\ep(z)$, then we can write many functions
of it\footnote{
We can define a function $w(\ep)$ of the distribution 
$\ep(z)=\lim\ep_n(z)$ by $w(\ep(z))=\lim w(\ep_n(z))$.}
still satisfying this property, 
\footnote{The possible appearance of the sign factors of this
type in the orbifold constructions was mentioned before;
see, e.g., Refs.~\cite{lama2,gero2}.
}
\bea
\ep(z), \quad \frac{1}{\ep(z)}\, , \quad
\frac{2\ep(z)}{1+\ep(z)^2}\, , \quad \text{etc.}
\eea
The $\ep(z)^2$ in Eq.~(\ref{epmis}) now changes as
\bea
\ep(z)^2 \qquad \longrightarrow \qquad 
\ep_5(z)\ep_6(z) \; .
\eea
Therefore, by choosing $\ep_5(z)=\ep(z)$ and $\ep_6(z)=1/\ep(z)$,
or vice versa, we eliminate the mismatch!

One should be careful, however, because such a modification can
change some of the relations used before. Namely, we
used $\ep^\prime(z)=2\da(z)$. It turns out, however, that we are safe 
since\footnote{
Note that although $\ep^\prime(z)=2\da(z)$ and $(1/\ep(z))^\prime=2\da(z)$
are both true in the distributional sence, the \emph{functional}
relation $\ep_n^\prime(z)=2\da_n(z)$ does not hold between
$1/\ep_n(z)$ and $\da_n(z)$.}
\bea
\left( \frac{1}{\ep(z)} \right)^\prime = -\frac{1}{\ep(z)^2}\ep^\prime(z)
=-2\frac{1}{\ep(z)^2}\da(z)=2\da(z) \; .
\eea
We used here the following relation,
\bea
\ep(z)^{-2}\da(z)=-\da(z) \; ,
\eea
which can be proven in the same way as Eq.~(\ref{ep2}). Namely,
\bea
\int_{-a}^{+a} \ep(z)^{-2}\da(z) dz 
= \frac{1}{2}\int_{-a}^{+a} \ep^{-2} d\ep
=-\frac{1}{2}\ep(z)^{-1}\Big|_{-a}^{+a}
=-\frac{1}{2}(1-(-1))=-1 \; .
\eea

We will find a more convincing proof of the necessity to introduce
different $\ep_i(z)$ for different fields when checking 
supersymmetry of our action in the orbifold picture. 

\subsection{Another addition to the $Y$-term}
We see now that the presence of the $e_5^\hf F^{n5}B_n$ term in
the boundary action follows most easily from the requirement that the
$Y$-term match singularities of the bulk Lagrangian.
We will now use this approach to find another term which should be
included in the $Y$-term.

In our expressions for the boundary terms of 
the general and supersymmetry
variations of the bulk action, we ignored a contribution from
the following term,
\bea
\label{Fff}
- i\frac{\sqrt6}{16}F_{MN}
\left(
2\Psitil^{Mi} \Psi_i^N + \Psitil_P^i \Ga^{MNPQ} \Psi_{Qi} 
\right) 
\in \mc{L}_5 \; .
\eea
The reason was that its contribution to the boundary term
of the supersymmetry variation (if it is at all non-zero) comes
from $\da_\Eta B_M$ and thus is \emph{quartic} in fermions,
which is of higher order than we consider. But its contribution
to the boundary term of the general variation is of quadratic
order in fermions and thus should be included.
(Note that the variation $\da_\Eta \Psi_{Mi}$ in Eq.~(\ref{Fff}) 
does not contribute to the boundary term of the
supersymmetry variation as the explicit calculation of
$\wt K_M$ in Eq.~(\ref{susytd}) shows \cite{thesis}.)

This part of $\mc{L}_5$ produces brane-localized terms 
because $F_{m5}\ni -2B_m^{(+)}\da(z)$ . Since the singular part
of $e_5^\hf\mc{L}_5$ should match onto $2Y^{(+)}\da(z)$, we find the
following contribution to the $Y$-term,
\bea
\label{Bff}
Y^{(+)} &\ni& \frac{\sx}{4}i\ga^{mk}B_k^{(+)}
(\psi_{m2}\psi_{\hf 1}-\psi_{m1}\psi_{\hf 2}) \nn\\
&&\qquad -\frac{\sx}{8}i\ep^{mnpq}B_m^{(+)}
(\psi_{p2}\si_n\psibar_{q2}+\psi_{p1}\si_n\psibar_{q1})+h.c. \; ,
\eea
where $\psi_{\hf 1,2}=e_\hf^5\psi_{51,2}+e_\hf^m\psi_{m1,2}$.
This contribution, however, does not change our previous
analysis. 

Indeed, because of the $e_5^\hf F^{n5}B_n$ term, the
gauge invariance of the bulk-plus-boundary action is broken
(on the boundary)
and we have to use the $B_m^{(+)}=0$ boundary condition.
The terms in Eq.~(\ref{Bff}) are, therefore, harmless
for the supersymmetry variation unless we vary $B_m$ itself, 
but this is of higher order in fermions. They do modify
the natural boundary conditions, making $B_m\;\sim\;\mc{O}(\psi^2)$,
but this is again of higher order in our approximation.

Also, by construction, these terms match singularities
of the bulk Lagrangian and thus do not appear in the 
bulk-plus-brane action of the orbifold picture.

\subsection{Result: the orbifold action}
We found that the $Y$-term of the boundary picture
matches (with some subtleties) onto the brane-localized terms 
arising from the singularities of the bulk Lagrangian $\mc{L}_5$. 
As a result, the total action (\ref{pmsum}) reduces to
\bea
S=\int_{\mc{M}_5} d^5x e_5 \mc{L}_5
+\int_\Si d^4x e_4 \mc{L}_4 \; ,
\eea
where $\mc{M}_5=\mathbb{R}^{1,4}$ is the (curved) 
space-time \emph{without} boundary,
with $z\in(-\infty, +\infty)$,
and $\Si$ denotes the brane at $z=0$.
The brane Lagrangian $\mc{L}_4$ is \emph{twice} the boundary
Lagrangian (not including the $Y$-term) evaluated on the $\mc{M}_+$
side of $\Si$,
\bea
\mc{L}_4 = 2\mc{L}_B^{(\al)}{}^{(+)} \; .
\eea

\section{Supersymmetry in the orbifold picture}
\label{sec-sg9}
Here we check explicitly whether the bulk-plus-brane
action, constructed starting from the boundary picture,
is in fact supersymmetric in the orbifold picture.
In the process, we find that using different 
$\ep_i(z)$ for different odd fields is essential and
that checking supersymmetry \emph{without} the use of the 
boundary conditions fixes the $\ep_i(z)$ \emph{uniquely}.

\subsection{Bulk-plus-brane action}
Our bulk-plus-brane action is
\bea
\label{bpbrac}
S=S_5+S_4 
= \int d^5x e_5\mc{L}_5 + \int d^5x e_4 \da(z)\mc{L}_4 \; ,
\eea
where $S_5$ is the bulk supergravity action (\ref{bulkaction}),
and $\mc{L}_4$ is the following brane Lagrangian,
\bea
\label{L4}
\mc{L}_4 =
-6\la_1+2\Big[
\al_{11}\psi_{m1}\si^{mn}\psi_{n1}
+2\al_{12}\psi_{m1}\si^{mn}\psi_{n2}^{(+)}
+\al_{22}\psi_{m2}^{(+)}\si^{mn}\psi_{n2}^{(+)}+h.c. \Big] \; .
\eea
The parameters are fixed in terms of $\la$, $\vec q$ and 
$\al\equiv\al^{(+)}$,
\bea
\al_{11}=\frac{-\al}{1+\al\al^\ast}, \qquad
\al_{12}=\frac{-\al\al^\ast}{1+\al\al^\ast}, \qquad
\al_{22}=\frac{\al^\ast}{1+\al\al^\ast}
\eea
\bea
\label{laalorb}
\la_1=-\frac{\al q_{12}^\ast+\al^\ast q_{12}
+(\al\al^\ast-1)q_3^{(+)} }
{1+\al\al^\ast}\la \; .
\eea
This bulk-plus-brane action was derived starting from 
the boundary picture. (Note that setting $\al=0$ kills
all the fermionic terms in $\mc{L}_4$. Therefore, in the
orbifold picture, the transition between the 
$\al=0$ and $\al\neq 0$ cases is not as straightforward
as in the boundary picture.)

We would now like to check explicitly that the action
is supersymmetric in the orbifold picture under the local
$N=2$ supersymmetry restricted on the brane by the boundary
condition 
\bea
\eta_2^{(+)}=\al\eta_1 \quad \text{on}\;\Si \; .
\eea
As we found in the previous section, it may be necessary to
use the freedom of defining different $\ep(z)$ for different fields.
We therefore set
\bea
\eta_2=\ep_1\eta_2^{(+)}, \quad
\psi_{m2}=\ep_2\psi_{m2}^{(+)}, \quad
K_{ma}=\ep_3 K_{ma}^{(+)}, \quad
q_3=\ep_4 q_3^{(+)}, \quad
B_m=\ep_5 B_m^{(+)} \; .
\eea
We will see that $\ep_i$ are either $\ep(z)$ or $1/\ep(z)$,
so that we can freely use $\ep_i^\prime(z)=2\da(z)$.

\subsection{Supersymmetry variation of the bulk action}
The supersymmetry variation of the bulk Lagrangian produces a
total derivative term, Eq.~(\ref{susytd}), 
which was important in the boundary
picture but integrates to zero on the orbifold. 
But on the orbifold we get
additional brane-localized contributions from the bulk action
due to the discontinuities in the fields and parameters.

First, we promoted the parameter $q_3$ to a function,
\bea
q_3(z)=\ep_4(z) q_3^{(+)} \; ,
\eea
where $q_3^{(+)}$ is a constant. Performing the supersymmetry
variation of $S_5$ without assuming the parameters to
be constant, we find \cite{bb1} (dropping the total derivative term),
\bea
\da_\Eta^{(1)} S_5 =\int d^5x e_5 \Big\{
-3i\Psitil_M^i\Ga^{MN}\Eta_j\p_N(\la\Qij)
-i\sx \Psitil_M^i\Ga^{MNK}\Eta_j B_K\p_N(\la\Qij) \Big\} \; .
\eea
In our case only $\p_5 q_3\neq 0$. 
Going into our gauge ($e_m^\hf=e_a^5=0$),
we obtain
\bea
\label{var1}
\da_\Eta^{(1)} S_5 &=& \int d^5x e_5 e_\hf^5 
\Big[ 2\la q_3^{(+)}\da(z) \Big] \Big\{
3i(-\psi_{m1}\si^m\etabar_1+\psi_{m2}\si^m\etabar_2) \nn\\
&&\hspace{100pt}
+2\sx i(\psi_{m2}\si^{mn}\eta_1+\psi_{m1}\si^{mn}\eta_2)B_n
+h.c. \Big\} \; ,
\eea
where we used $q_3{}^\prime(z)=2 q_3^{(+)}\da(z)$.

Second, because we have $\p_5$ hitting odd fields in our 
supersymmetry transformations, the transformations are 
\emph{singular} and thus \emph{not well-defined on the brane}. 
In our analysis of the Mirabelli and Peskin model \cite{my1}, 
we showed that in order for the (on-shell) supersymmetry
algebra to close onto the (singular) orbifold equations of motion
we need to modify the supersymmetry transformations by adding
appropriate $\da(z)$-terms. The modifications should be such
that when the (natural) boundary conditions are taken into 
account the supersymmetry transformations 
\emph{become non-singular}
on the brane. This approach was already used in Ref.~\cite{bb1}.

By inspection of our supersymmetry transformations, we see that 
$\p_5$ hits $\eta_2$ in $\da\psi_{52}$. Therefore, we
modify the supersymmetry transformations by \emph{adding} to
$\da\psi_{52}$ a new piece,
\bea
\label{mod52}
\da_\Eta^{(2)}\psi_{52}=-4\eta_2^{(+)}\da(z)=-4\al\eta_1\da(z) \; ,
\eea
which subtracts the singular piece in $\da\psi_{52}$.
This modification produces an additional brane-localized
contribution to the supersymmetry variation of the bulk action,
\bea
\da_\Eta^{(2)}S_5 &=& \int d^5x e_5 e_\hf^5 [-4\da(z)]\Big\{ 
-2\psi_{m1}\si^{mn}\wh D_n\eta_2^{(+)}
+i K_{na}\psi_{m2}\si^{mn}\si^a\etabar_2^{(+)} \nn\\
&&-\frac{3}{2}\la i
(q_3\psi_{m2}+q_{12}\psi_{m1})\si^m\etabar_2^{(+)}
+\frac{\sx}{4}i\ga^{mk}F_{k\hf}\psi_{m1}\eta_2^{(+)} +h.c.
\Big\} \; ,
\eea
where $\ga^{mn}=e^{ma}e_a^n$ and
$F_{m\hf}=e_\hf^5 F_{m5}+e_\hf^n F_{mn}$. (The $F_{mn}$ terms
not appearing in $F_{m\hf}$, as well as the $B_m$ terms, are
not shown here.)

All other $\p_5$ in the supersymmetry transformations appear
only via $F_{m5}$. However, as we will explain in detail later,
when the natural boundary condition on $B_m$ is
$B_m^{(+)}=0$ on $\Si$, no further 
modifications to the supersymmetry transformations are necessary.

We will return to the discussion of the $B_m$ and $F_{mn}$ terms
in the next section. For now we simply
set $B_m=0$ and $F_{mn}=0$. (But we will keep $F_{m5}$.)

\subsection{Supersymmetry variation of the brane action}
The supersymmetry variation of the brane action gives
\bea
\da_\Eta S_4 &=& \int d^5x e_4 \da(z) \Big\{
-6\la_1 e_a^m\da_\Eta e_m^a +4\Big[
(\al_{11}\psi_{m1}+\al_{12}\psi_{m2}^{(+)})
\si^{mn}\da_\Eta\psi_{n1} \nn\\[5pt]
&&\hspace{100pt}
+(\al_{22}\psi_{m2}^{(+)}+\al_{12}\psi_{m1})
\si^{mn}\da_\Eta\psi_{n2}^{(+)} +h.c.\Big] \Big\} \; .
\eea
The (induced on the brane) supersymmetry transformations are
\bea
\da_\Eta e_m^a &=& 
-i(\psi_{m1}\si^a\etabar_1+
\ep_1\ep_2\psi_{m2}^{(+)}\si^a\etabar_2^{(+)})+h.c. \nn\\[5pt]
\da_\Eta\psi_{m1} &=& 2\wh D_m\eta_1
+i\ep_1\ep_3 K_{ma}^{(+)}\etabar_2^{(+)}
+i\la\si_m(q_{12}^\ast\etabar_1
+\ep_1\ep_4 q_3^{(+)}\etabar_2^{(+)}) \nn\\
&&\hspace{150pt}
-\frac{4i}{2\sx}(\si_m{}^n+\da_m^n)\eta_1 F_{n\hf} \nn\\
\da_\Eta\psi_{m2}^{(+)} &=& 2\wh D_m\eta_2^{(+)}
-i K_{ma}^{(+)}\etabar_1
+i\la\si_m(q_3^{(+)}\etabar_1-q_{12}\etabar_2^{(+)})\nn\\
&&\hspace{150pt}
-\frac{4i}{2\sx}(\si_m{}^n+\da_m^n)\eta_2^{(+)} F_{n\hf} \; ,
\eea
where the $\ep_i(z)$ factors have been explicitly shown.
(Note that the variations of the even fields contain products of 
$\ep_i$\;,
whereas the variation of $\psi_{m2}^{(+)}$ contains no $\ep_i$
and simply corresponds to evaluating the bulk transformation 
of the odd field on the positive side of the brane.)

\subsection{Supersymmetry variation of the total action}
Writing all the $\ep_i(z)$ factors explicitly in the expressions
for the supersymmetry variation of the bulk action, we find
\bea
\da_\Eta^{(1)} S_5 = \int d^5x e_4 \da(z) \Big\{
6i\la q_3^{(+)} \Big( -\psi_{m1}\si^m\etabar_1
+\ep_1\ep_2\psi_{m2}^{(+)}\si^m\etabar_2^{(+)} \Big) 
+h.c. \Big\}
\eea
and
\bea
\da_\Eta^{(2)}S_5 &=& \int d^5x e_4 \da(z)\Big\{ 
8\psi_{m1}\si^{mn}\wh D_n\eta_2^{(+)}
-4i \ep_2\ep_3 K_{na}^{(+)}\psi_{m2}^{(+)}
\si^{mn}\si^a\etabar_2^{(+)} \nn\\
&&\hspace{80pt}
+6\la i
\Big( \ep_2\ep_4 q_3^{(+)}\psi_{m2}^{(+)}+q_{12}\psi_{m1} \Big)
\si^m\etabar_2^{(+)} \nn\\[5pt]
&&\hspace{80pt}
-\sx i\ga^{mk}F_{k\hf}\psi_{m1}\eta_2^{(+)} +h.c.
\Big\} \; .
\eea
Adding all three contributions and setting $\eta_2^{(+)}=\al\eta_1$,
we obtain the following expression for the supersymmetry variation
of our bulk-plus-brane action,
\bea
\da_\Eta S &=& \da_\Eta^{(1)}S_5+\da_\Eta^{(2)}S_5+\da_\Eta S_4 
= \int d^5x e_4 \da(z)\Big\{ 
8\Big[ \wt C_1\psi_{m1}+\wt C_2\psi_{m2}^{(+)} \Big]
\si^{mn}\wh D_n\eta_1  \nn\\[5pt]
&& 
+4i K_{na}^{(+)}\Big[ \wt C_3\psi_{m1}+\wt C_4\psi_{m2}^{(+)} \Big]
\si^{mn}\si^a\etabar_1
-6i \Big[ \wt C_5\psi_{m1}+\wt C_6\psi_{m2}^{(+)} \Big]
\si^m\etabar_1 \nn\\[5pt]
&&\hspace{40pt}
-\sx i\ga^{mk}F_{k\hf}
\Big[ \wt C_1\psi_{m1}+\wt C_2\psi_{m2}^{(+)} \Big]\eta_1
+h.c.\Big\} \; ,
\eea
where the coefficients are
\bea
\wt C_1 &=& \al_{11}+\al(\al_{12}+1) \nn\\
\wt C_2 &=& \al_{12}+\al\al_{22} \nn\\
\wt C_3 &=& \ep_1\ep_3\al_{11}\al^\ast-\al_{12}  \nn\\
\wt C_4 &=& \al^\ast(\ep_1\ep_3\al_{12}-\ep_2\ep_3)-\al_{22} \nn\\
\wt C_5 &=& -\la_1+\la \wt A_1 \nn\\
\wt C_6 &=& -\ep_1\ep_2\la_1\al^\ast+\la \wt A_2 
\eea
with
\bea
\wt A_1 &=& \al_{11}(\ep_1\ep_4 q_3^{(+)}\al^\ast+q_{12}^\ast)
+(\al_{12}+1)(q_3^{(+)}-q_{12}\al^\ast) \nn\\[5pt]
\wt A_2 &=& \al_{12}(\ep_1\ep_4 q_3^{(+)}\al^\ast+q_{12}^\ast)
+\al_{22}(q_3^{(+)}-q_{12}\al^\ast) 
-(\ep_1\ep_2+\ep_2\ep_4)q_3^{(+)}\al^\ast \; .
\eea
The total action is supersymmetric (subject only to the
$\eta_2=\al\eta_1$ and $B_m=0$ boundary conditions) if 
all $\wt C_i$ vanish. Comparing $\wt C_i$ with $C_i$ in
Eq.~(\ref{Cs}), we see that this happens if and only if
\bea
\ep_1\ep_3=1, \quad
\ep_2\ep_3=-1, \quad
\ep_1\ep_2=1, \quad
\ep_1\ep_4=1, \quad
\ep_1\ep_2+\ep_2\ep_4=0 \; .
\eea
But this must be true \emph{when multiplied by} $\da(z)$!
Since we know that
\bea
\ep^2\da(z)=\frac{1}{3}\da(z), \qquad
\ep^{-2}\da(z)=-\da(z) \; ,
\eea
we see that our bulk-plus-brane action is supersymmetric
provided we choose
\bea
\label{neweps}
\ep_1=\ep(z), \quad \ep_2=\ep_3=\ep_4=\frac{1}{\ep(z)} \; .
\eea

\subsection{Connection with earlier work}
Since this assignment differs from 
\bea
\label{oldeps}
\ep_1=\ep_2=\ep_3=\ep_4=\ep(z) \; ,
\eea
assumed in Ref.~\cite{bb1}, let us reproduce that calculation in which
the boundary conditions
\bea
K_{ma}^{(+)}=\la e_{ma}, \qquad \psi_{m2}^{(+)}=\al\psi_{m1}
\quad \text{on}\; \Si
\eea
were used in checking supersymmetry of the bulk-plus-brane action.

If we use these boundary conditions in the supersymmetry 
variation of the action, we find
\bea
\da_\Eta S &=& \int d^5x e_4 \da(z)\Big\{ 
(\wt C_1+\al \wt C_2)\Big[ 8\psi_{m1}\si^{mn}\wh D_n\eta_1
-\sx i\ga^{mk}F_{k\hf}\psi_{m1}\eta_1 \Big] \nn\\[5pt]
&&\hspace{100pt}
+6i \wt M\psi_{m1}\si^m\etabar_1 +h.c. \Big\} \; ,
\eea
where 
\bea
\wt M=-\la_1(\wt C_3+\al \wt C_4)-(\wt C_5+\la \wt C_6) \; .
\eea
We already saw in Eq.~(\ref{shortLB})
that using the $\psi_{m2}$ boundary condition
in the brane action reduces the coefficients $\al_{ij}$ as
follows,
\bea
(\al_{11}, \al_{12}, \al_{22}) \quad \longrightarrow \quad
(-\al, 0, 0) \; .
\eea
This makes
\bea
&&\wt C_1=\wt C_2=0, \qquad
\wt C_3=-\ep_1\ep_3\al\al^\ast, \qquad
\wt C_4=-\ep_2\ep_3\al^\ast \nn\\[3pt]
&&\wt C_5=-\la_1-\la\al(\ep_1\ep_4 q_3\al^\ast+q_{12}^\ast)
+\la(q_3-q_{12}\al^\ast) \nn\\[3pt]
&&\wt C_6=-\ep_1\ep_2\la_1\al^\ast
-\la(\ep_1\ep_2+\ep_2\ep_4)q_3\al^\ast
\eea
and, therefore,
\bea
\wt M &=& \la_1\Big[ 
1+(\ep_1\ep_2+\ep_1\ep_3+\ep_2\ep_3)\al\al^\ast \Big]
\nn\\[5pt]
&&\qquad
+\la \Big\{q_{12}\al^\ast +q_{12}^\ast \al +q_3\big[
(\ep_1\ep_4+\ep_1\ep_2+\ep_2\ep_4)\al\al^\ast-1 \big] \Big\} \; .
\eea
The action is supersymmetric provided $\wt M\da(z)=0$, which
is equivalent to
\bea
&(\ep_1\ep_2+\ep_1\ep_3+\ep_2\ep_3)\da(z)=\da(z)& \\[5pt]
&(\ep_1\ep_4+\ep_1\ep_2+\ep_2\ep_4)\da(z)=\da(z)&,
\eea
when Eq.~(\ref{laalorb}) is taken into account. We see
that \emph{both} choices of $\ep_i(z)$, Eqs.~(\ref{neweps})
and (\ref{oldeps}), satisfy these conditions!

We, therefore, conclude that checking supersymmetry with the
boundary conditions taken into account is \emph{insufficient} 
to distinguish between the different $\ep_i(z)$ assignments.
Without the use of the boundary conditions, supersymmetry of the
bulk-plus-brane action provides more consistency checks and
requires the assignment in Eq.~(\ref{neweps}). Namely,
\bea
\eta_2=\ep(z)\eta_2^{(+)}, \quad
\psi_{m2}=\frac{1}{\ep(z)} \psi_{m2}^{(+)}, \quad
K_{ma}=\frac{1}{\ep(z)} K_{ma}^{(+)}, \quad
q_3=\frac{1}{\ep(z)} q_3^{(+)} \; .
\eea

\section{Fate of $B_m$ terms on the orbifold}
\label{sec-sg10}
In this section we will write down the $B_m$ and $F_{mn}$
terms appearing in the supersymmetry variation of the
bulk-plus-brane action explicitly. We will find that they
generally do not cancel so that the use of the $B_m^{(+)}=0$
boundary condition appears to be necessary for supersymmetry.

\subsection{$F_{mn}$ terms in the supersymmetry variation}
We consider first the $F_{mn}$ terms. Those that appear
in the supersymmetry transformations and in the bulk action
in the combination 
$F_{m\hf}=e_\hf^5 F_{m5}+e_\hf^n F_{mn}$,
go through the supersymmetry variation in this combination
and cancel just as the $F_{m5}$ terms we considered in 
the previous section. We will, therefore, omit them here.
Among the remaining $F_{mn}$ terms, we need only the following terms
in the bulk Lagrangian,
\bea
\mc{L}_5=i\frac{\sx}{8}\ep^{mnpq}F_{pq}
(\psi_{m2}\si_n\psibar_{52})e_\hf^5 +h.c. \; ,
\eea
and only the following terms in the supersymmetry transformations,
\bea
&& \da_\Eta \psi_{m1}=+\frac{1}{2\sx}(i\eps_m{}^{nkl}\si_l
+4\da_m^n\si^k)\etabar_{2} F_{nk} \nn\\
&& \da_\Eta \psi_{m2}=-\frac{1}{2\sx}(i\eps_m{}^{nkl}\si_l
+4\da_m^n\si^k)\etabar_{1} F_{nk} \; .
\eea
There are no ``$q_3\cdot F_{MN}$'' terms, thus 
$\da_\Eta^{(1)}S_5=0$.
The modification 
$\da_\Eta^{(2)}\psi_{52}=-4\eta_2^{(+)}\da(z)$ gives
\bea
\da_\Eta^{(2)} S_5=\int d^5x e_4\da(z)\Big[
-i\frac{\sx}{2}\ep^{mnpq}F_{pq}(\psi_{m2}\si_n\etabar_2^{(+)})
+h.c. \Big] \; .
\eea
Using a $\si$-matrix identity ($\ga^{mn}\equiv e_a^m e^{na}$),
\bea
\si^{mp}(i\eps_p{}^{nkl}\si_l +4\da_p^n\si^k)
=\frac{3}{2}i\eps^{mnkl}\si_l+(\ga^{km}\si^n-\ga^{kn}\si^m) \; ,
\eea
we find the following expression for the supersymmetry
variation of the brane action,
\bea
\da_\Eta S_4 &=& \int d^5x e_4\da(z) \frac{4}{2\sx}\Big\{
\big(\al_{11}\psi_{m1}+\al_{12}\psi_{m2}^{(+)}\big)
\Big(
\frac{3}{2}i\eps^{mnkl}\si_l+\ga^{km}\si^n
\Big)\etabar_2
F_{nk} \nn\\
&&\hspace{20pt}
-\big(\al_{12}\psi_{m1}+\al_{22}\psi_{m2}^{(+)}\big)
\Big(
\frac{3}{2}i\eps^{mnkl}\si_l+\ga^{km}\si^n
\Big)\etabar_1
F_{nk}^{(+)} \Big\}+h.c.
\eea
Employing the $\ep_i(z)$ assignments and using the 
$\eta_2^{(+)}=\al\eta_1$ boundary condition, we obtain
the following expression for the supersymmetry variation
of our bulk-plus-brane action (showing only the $F_{mn}$ terms),
\bea
\da_\Eta S &=& \int d^5x e_4\da(z) \frac{4}{2\sx} F_{nk}^{(+)} 
\Big\{
\frac{3}{2}i\eps^{mnkl}
\big[Z_1\psi_{m1}+Z_2\psi_{m2}^{(+)}\big] \si_l\etabar_1 
\nn\\[5pt]
&&\hspace{127pt}
+\ga^{km}
\big[Z_3\psi_{m1}+Z_4\psi_{m2}^{(+)}\big] \si^n\etabar_1 
\Big\}+h.c. \; ,
\eea
where
\bea
\ba{l@{\qquad}l}
Z_1=\ep_1\ep_5\al_{11}\al^\ast-\al_{12}, &
Z_2=\ep_1\ep_5\al_{12}\al^\ast-\al_{22}-\ep_2\ep_5\al^\ast
 \\[5pt]
Z_3=\ep_1\ep_5\al_{11}\al^\ast-\al_{12}, &
Z_4=\ep_1\ep_5\al_{12}\al^\ast-\al_{22} \; .
\ea
\eea
We find that we can make $Z_1=Z_2=Z_3=0$ by choosing
\bea
\label{neweps2}
\ep_5=\frac{1}{\ep(z)} \qquad \Longleftrightarrow \qquad
B_m=\frac{1}{\ep(z)} B_m^{(+)} \; .
\eea
But we are still left with $Z_4=-\al^\ast$.
In order to cancel the remaining piece in the supersymmetry
variation,
\bea
\da_\Eta S = \int d^5x e_4\da(z) \Big\{
-\frac{4}{2\sx} \al^\ast \ga^{km}F_{nk}^{(+)}
\big( \psi_{m2}^{(+)}\si^n\etabar_1 \big) \Big\}+h.c. \; ,
\eea
we have to use the boundary condition $B_m^{(+)}=0$ on $\Si$.
We note, however, that we need it here
only in its (seemingly) gauge invariant form:
$F_{nk}^{(+)}=0$ on $\Si$.

\subsection{$B_m$ terms in the supersymmetry variation}
Repeating the above steps, we find that the $B_m$ terms 
give the following contributions,
\bea
\da_\Eta^{(1)} S_5 &=& \int d^5x e_4\da(z) \Big\{
4\sx i\la q_3^{(+)}(\psi_{m2}\si^{mn}\eta_1
+\psi_{m1}\si^{mn}\eta_2)B_n +h.c.\}
\\
\da_\Eta^{(2)} S_5 &=& \int d^5x e_4\da(z) \Big\{
4\sx i\la B_n
\big[ q_3 \psi_{m1}-q_{12}^\ast\big ] 
\psi_{m1}\si^{mn}\eta_2^{(+)}
+h.c. \Big\} 
\\
\da_\Eta S_4 &=& \int d^5x e_4\da(z) 4\sx i\la \Big\{
\big(\al_{11}\psi_{m1}+\al_{12}\psi_{m2}^{(+)}\big)
\si^{mn}(q_{12}^\ast\eta_2-q_3\eta_1) B_n
\nn\\
&&\hspace{20pt}
-\big(\al_{12}\psi_{m1}+\al_{22}\psi_{m2}^{(+)}\big)
\si^{mn}(q_3^{(+)}\eta_2^{(+)}+q_{12}\eta_1) B_n^{(+)}
\Big\}+h.c.
\eea
The total contribution to the supersymmetry variation of our
bulk-plus-brane action is the sum of these three,
\bea
\da_\Eta S=\int d^5x e_4\da(z) 
4\sx i\la B_n^{(+)}
\Big\{
W_1 \psi_{m1}\si^{mn}\eta_1
+W_2 \psi_{m2}^{(+)}\si^{mn}\eta_1
+h.c. \Big\} \; ,
\eea
where
\bea
W_1 &=& \al\al_{11}q_{12}^\ast +\al_{12}q_{12} +
q_3^{(+)}\big[
\ep_1\ep_5\al+\ep_4\ep_5\al-\ep_4\ep_5\al_{11}+\al\al_{12} \big]
\nn\\[5pt]
W_2 &=& \al\al_{12}q_{12}^\ast +\al_{22}q_{12} +
q_3^{(+)}\big[
\ep_2\ep_5-\ep_4\ep_5\al_{12}+\al\al_{22} \big] \; .
\eea
With our $\ep_i(z)$ assignments,
\bea
\ep_1=\ep(z), \qquad 
\ep_2=\ep_3=\ep_4=\ep_5=\frac{1}{\ep(z)} \; ,
\eea
the coefficients simplify to
\bea
W_1 &=& \al\al_{11}q_{12}^\ast +\al_{12}q_{12} -\al q_3^{(+)}
\nn\\[5pt]
W_2 &=& \al\al_{12}q_{12}^\ast +\al_{22}q_{12} - q_3^{(+)}
\eea
and can also be rewritten as 
\bea
W_1 &=& \frac{-\al}{1+\al\al^\ast}\Big[
(\al q_{12}^\ast+q_3)+\al^\ast(q_{12}+\al q_3) \Big]
\nn\\[5pt]
W_2 &=& \frac{1}{1+\al\al^\ast}\Big[
(\al^\ast q_{12}-q_3)-\al\al^\ast(\al q_{12}^\ast+q_3) \Big] \; .
\eea
These coefficients do not vanish unless
$\al=0$ and $q_3^{(+)}=0$ (or $\la=0$ for any $\al$).
In a general case, we need to use the 
$B_m^{(+)}=0$ boundary condition to cancel this part of
the supersymmetry variation.

This completes our check of supersymmetry of the
bulk-plus-brane action, Eq.~(\ref{bpbrac}). (We remind that
we work only to quadratic order in fermions.)
We found that besides the boundary condition on the supersymmetry
parameter, $\eta_2^{(+)}=\al\eta_1$, we need to use only
one other boundary condition: $B_m^{(+)}=0$.

In order to understand if the use of this boundary condition is
forced on us by the supersymmetry algebra, we have to
understand when the $U(1)$ gauge invariance is broken.
We discuss this in the next section.

\section{Gauge transformations on the orbifold}
\label{sec-sg11}
In this section we discuss the breaking of the general 
coordinate and the $U(1)$ gauge invariances on the orbifold.
We come to the conclusion that in order to reproduce the
results of the boundary picture discussion, we need to modify
all the gauge transformations on the orbifold
by making them \emph{non-singular}!
We show how the supersymmetry transformations are modified
for a general $B_m$ boundary condition, and find that the
algebra of the modified transformations closes as in the
boundary picture.

\subsection{Breaking of the general coordinate invariance}
The variation of the bulk Lagrangian
under the general coordinate transformation is a total
derivative,
\bea
\da_v \mc{L}_5=D_M (v^M\mc{L}_5) \; .
\eea
In the boundary picture, this produces a boundary term,
Eq.~(\ref{S5gct}),
the vanishing of which requires $v^5=0$ on the boundary,
and in turn leads to the restriction $\eta_2=\al\eta_1$
on the supersymmetry parameters. But on the orbifold the
total derivative integrates to zero! And it appears that
the restriction $(v^5)^{(+)}=0$ on the brane does not
arise.

If we take the point of view that the orbifold picture 
should reproduce all the major results of the boundary
picture (such as the breaking of a gauge invariance),
we are forced to make some modifications.

The necessary modification comes naturally from the 
requirement that the gauge transformations be the same
in the both pictures both in the bulk \emph{and} on the
brane/boundary. But if we take the transformations of the
boundary picture and assume them to be \emph{literally} the same
on the orbifold, we find that they are in general
\emph{singular} (and thus not well-defined) on the brane!
Indeed,
\bea
\da_v B_m &=& v^n\p_n B_m+v^5\p_5 B_m+B_n\p_m v^n+B_5\p_m v^5
=2 v^5 B_m^{(+)}\da(z)+\dots \nn\\[5pt]
\da_v B_5 &=& v^n\p_n B_5+v^5\p_5 B_5+B_n\p_5 v^n+B_5\p_5 v^5
=2 (v^5)^{(+)}B_5\da(z)+\dots \; ,
\eea
where the dots represent non-singular terms.
(We used here the fact that on the orbifold $v^5$ is odd,
whereas $v^m$ is even.) In order for the transformations
in the both pictures to agree on the boundary, we have to
\emph{modify} the transformations for the orbifold picture
by subtracting the singular pieces! For the $B_M$ field,
the modified general coordinate transformations are
as follows,
\bea
\label{modgct}
\da_v^\prime B_m &=& \da_v B_m -2 v^5 B_m^{(+)}\da(z) \nn\\[5pt]
\da_v^\prime B_5 &=& \da_v B_5 -2 (v^5)^{(+)}B_5\da(z) \; .
\eea
The transformations for other fields are appropriately modified.
It is clear, that the variation of the bulk Lagrangian under
the modified general coordinate transformation produces now
additional brane-localized terms which vanish only when 
$(v^5)^{(+)}=0$ on the brane. Therefore, all the related
conclusions of the boundary picture are now reproduced.

\subsection{Breaking of the $U(1)$ gauge invariance}
In the boundary picture, the variation of the bulk Lagrangian 
under the $U(1)$ gauge transformation is a total derivative
(arising from the Chern-Simons term),
\bea
\da_u \mc{L}_5 = D_K \Big[
-u \frac{1}{6\sqrt6}\;\eps^{MNPQK} F_{MN}F_{PQ} \Big]  \; .
\eea
It produces a boundary term,
Eq.~(\ref{S5gbp}), which tells us that the bulk action is gauge
invariant only if some boundary condition is imposed. Namely,
$u=0$ or $F_{mn}=0$ on the boundary.

On the orbifold, the total derivative is also generated,
but it integrates to zero.
It turns out, however, 
that under the original $U(1)$ gauge transformation,
Eq.~(\ref{U1tr}),
\bea
\da_u B_M =\p_M u , \qquad
\da_u \Psi_{Mi}= u \frac{\sx}{2}\la\Qij\Psi_{Mj} \; ,
\eea
we now do get brane-localized terms in the variation
of the bulk Lagrangian.

The variation receives a new contribution on the orbifold
because the modified covariant derivative in 
Eq.~(\ref{mcovder}) is \emph{not} covariant under the $U(1)$
gauge transformation when the parameters are not constant,
\bea
\label{laq3term}
\da_u (\wt D_M\Psi_{Ni})
=u \frac{\sx}{2}\la\Qij(\wt D_M \Psi_{Nj})
+u \frac{\sx}{2}\p_M(\la\Qij)\Psi_{Nj} \; .
\eea
The variation of the Lagrangian relevant in the orbifold
picture is, therefore,
\bea
\da_u \mc{L}_5=\frac{i}{2}\Psitil_M^i\Ga^{MNK}
\Big[
u \frac{\sx}{2}\p_N(\la\Qij)\Psi_{Kj} \Big] \; .
\eea
Since only $q_3$ is not a constant, we obtain 
\bea
\label{gvarS5orb}
\da_u S_5= \int d^5x e_4 \da(z) \frac{\sx}{2}\la u \big[
-4i q_3^{(+)}(\psi_{m1}\si^{mn}\psi_{n2})+h.c. \big] \; .
\eea
Therefore, the bulk action is not gauge invariant if 
$\la q_3^{(+)}\neq 0$. 

Our brane Lagrangian is $\mc{L}_4=2 \mc{L}_B^{(\al)}{}^{(+)}$, 
where $\mc{L}_B^{(\al)}$ is the boundary Lagrangian of the
boundary picture \emph{without} the $Y$-term. Its variation
under the $U(1)$ gauge transformation is given in 
Eq.~(\ref{gvarLBal}). Therefore, the brane Lagrangian is
by itself gauge invariant only when $\al=0$ (so that the
Lagrangian vanishes) or when $\la=0$ (so that the gauge
transformation does not act on the fermions).

It is easy to check that the bulk-plus-brane action
is gauge invariant only when the bulk and the
brane actions are \emph{separately} gauge invariant.
(That is the sum of the two contributions still vanishes
only when $\al=0$ and $q_3^{(+)}=0$. Or when $\la=0$.)

On the other hand, the boundary condition 
$\eta_2^{(+)}=\al\eta_1$ is gauge invariant when
\bea
\la \Big[
\big(q_{12}+\al q_3^{(+)}\big)
-\al\big(\al q_{12}^\ast-q_3^{(+)}\big)
\Big] =0 \; .
\eea
It follows that the $U(1)$ gauge invariance in the orbifold
picture is broken by either the bulk-plus-brane action 
or the fermionic boundary condition unless $\la=0$ ! 

Since this is drastically different from the way the
$U(1)$ gauge invariance is broken in the boundary picture,
we have to make some modifications if we would like the
two pictures to describe the same physics.

\subsection{Modified $U(1)$ gauge transformation}
We found that taking the $U(1)$ gauge transformation in
the orbifold picture
to be \emph{literally} the same as in the boundary
picture leads to very different conclusions about the
breaking of the gauge invariance in the both pictures.
Therefore, as in the case of the general coordinate invariance,
we are led to modify the $U(1)$ gauge transformation
in the orbifold picture. The modification affects
only $B_5$,
\bea
\label{modU1}
\da_u^\prime B_5 =\p_5 u -2u^{(+)}\da(z) \; .
\eea
(The parameter $u$ is odd.) The modified transformation
is \emph{non-singular} on the brane and coincides with
the $U(1)$ transformation induced on the boundary in the
boundary picture,
\bea
\da_u B_5 =\p_5 u \quad \text{on}\;\p\mc{M}
\qquad \Longleftrightarrow \qquad
\da_u^\prime B_5 =\p_5 u^{(+)} \quad \text{on}\;\Si \; .
\eea

The variation of the bulk action under the modified
$U(1)$ gauge transformation produces the following
brane-localized term (from the variation of the Chern-Simons term
in the bulk Lagrangian),
\bea
\label{CScontr}
\da^\prime_u S_5 \ni \int d^5x e_4 \da(z) \left\{
u^{(+)}\frac{2}{6\sx}\eps^{mnpq}F_{mn}F_{pq} \right\} \; ,
\eea
which is the orbifold version of Eq.~(\ref{S5gbp}).

For the covariant derivative defined in 
Eq.~(\ref{mcovder}) we now obtain, instead of Eq.~(\ref{laq3term}),
\bea
\da_u (\wt D_5\Psi_{Ni})
&=& u \frac{\sx}{2}\la\Qij(\wt D_5 \Psi_{Nj}) \nn\\[5pt]
&&+u \frac{\sx}{2}\p_5(\la\Qij)\Psi_{Nj} 
+\sx u^{(+)}\da(z)(\la\Qij)\Psi_{Nj} \; .
\eea
The variation of the the fermionic part of the
bulk action now gets two contributions.
(Let us use the shorthand notation 
$\psi_i\psi_j\equiv\psi_{mi}\si^{mn}\psi_{nj}$
for the following.)
The first contribution, arising from the jumping parameter
$q_3=\ep_4 q_3^{(+)}$, is
\bea
\da_u^\prime{}^{(1)} S_5 = \int d^5x e_4 \da(z) 
\frac{\sx}{2}\la u
\Big[
-4i q_3^{(+)}\psi_1\psi_2 +h.c.
\Big] \; ,
\eea
and the second one, arising from the extra piece 
$\da_u^\prime{}^{(2)}B_5 =-2u^{(+)}\da(z)$ in the modified gauge
transformation, is
\bea
\da_u^\prime{}^{(2)} S_5 = \int d^5x e_4 \da(z) 
\frac{\sx}{2}\la u^{(+)}
\Big[
-4i q_3\psi_1\psi_2
-2i q_{12}(\psi_1\psi_1
+\psibar_2\psibar_2) +h.c.
\Big]  \; .
\eea
The variation of the brane action is \emph{twice} that in 
Eq.~(\ref{gvarLBal}), with $\psi_{m2}$ and $q_3$ evaluated
on the $\mc{M}_+$ side of $\Si$. Namely,
\bea
\da_u^\prime S_4 = \int d^5x e_4 \da(z) 
\frac{\sx i\la u^{(+)}}{1+\al\al^\ast}
\Big[
X_1\psi_1\psi_1 +X_2\psi_1\psi_2^{(+)} +X_3\psi_2^{(+)}\psi_2^{(+)}
\Big]+h.c. \; ,
\eea
where
\bea
X_1=X_3^\ast=2\al\big( \al^\ast q_{12}-q_3^{(+)} \big), \qquad
X_2=2(\al q_{12}^\ast-\al^\ast q_{12}) \; .
\eea
The total variation of the fermionic part 
of the bulk-plus-brane action under
the (modified) $U(1)$ gauge transformation is the sum
of the three contributions,
\bea
\da_u^\prime S \ni \int d^5x e_4 \da(z) 
\frac{\sx i\la u^{(+)}}{1+\al\al^\ast}
\Big[
\wt X_1\psi_1\psi_1 +\wt X_2\psi_1\psi_2^{(+)} 
+\wt X_3\psi_2^{(+)}\psi_2^{(+)}
\Big]+h.c. \; ,
\eea
where
\bea
\wt X_1 &=& X_1-q_{12}(1+\al\al^\ast) \nn\\
\wt X_3 &=& X_3+ q_{12}^\ast(1+\al\al^\ast)\ep_2\ep_2 \nn\\
\wt X_2 &=& X_2-2 q_3^{(+)}(1+\al\al^\ast)[\ep_u\ep_2+\ep_2\ep_4]
\eea
after all the $\ep_i(z)$ factors are separated. The $\ep_u$
is such a factor for the odd parameter $u$,
\bea
u(x,z)=\ep_u(z) u^{(+)}(x, |z|) \; .
\eea 
With our $\ep_i(z)$ assignments, Eq.~(\ref{neweps}), we have
$\ep_2\ep_2\da(z)=-\da(z)$. Therefore,
\bea
\wt X_1=\wt X_3^\ast 
=\al(\al^\ast-q_3^{(+)})-(\al q_3^{(+)}+q_{12}) \; ,
\eea
which is exactly the coefficient $W$ in Eq.~(\ref{gvarLB})!
More than that, if we choose 
\bea
\label{neweps3}
\ep_u=\ep(z) \; ,
\eea
we get $[\ep_u\ep_2+\ep_2\ep_4]\da(z)=0$, so that the
equation (\ref{gvarLB}) is reproduced completely!
This means that the fermionic part of
our bulk-plus-brane action is now gauge
invariant with the same restriction on the parameters $\vec q$
and $\al$, Eq.~(\ref{ginval}), as is necessary for the gauge
invariance of the boundary condition $\eta_2^{(+)}=\al\eta_1$ !

After this choice of $\vec q$ and $\al$ is made, the variation
of the bulk-plus-brane action under the 
$U(1)$ gauge transformation has only one uncanceled piece,
Eq.~(\ref{CScontr}),
\bea
\label{CScontr2}
\da^\prime_u S = \int d^5x e_4 \da(z) \left\{
u^{(+)}\frac{2}{6\sx}\eps^{mnpq}F_{mn}F_{pq} \right\} \; ,
\eea
which is the orbifold picture analog of Eq.~(\ref{S5gbp}).
Therefore, our modification of the gauge transformations
in the orbifold picture, Eq.~(\ref{modU1}), leads to agreement
with conclusions of the boundary picture.

\subsection{Modified supersymmetry transformations}
\label{sec-modsusy}
The modification of the supersymmetry transformation for
$\psi_{52}$, Eq.~(\ref{mod52}), now becomes just a part
of the general modification of \emph{all} the gauge transformations
in the orbifold picture. Indeed,
\bea
\da_\Eta^\prime \psi_{52}=\da_\Eta \psi_{52}-4\eta_2^{(+)}\da(z)
\eea
is an analog of Eq.~(\ref{modU1}) which makes the supersymmetry
transformations \emph{non-singular} on the brane, so that the
induced on the brane transformations are exactly the same as
those in the boundary picture.

The supersymmetry transformations should also be modified
when the boundary condition on $B_m$ is no longer $B_m^{(+)}=0$,
but instead $B_m^{(+)}=J_m$. This happens, for example,
when one couples the
$B_m$ field to some brane-localized matter.
The $J_m$ is then a composite of the brane matter fields.
(The coupling of brane-localized matter to the bulk supergravity
in five dimensions is discussed in 
Refs.~\cite{zu3,matter1,matter2,matter3,bb3,matter4}.)

The necessary modifications in the supersymmetry transformations
can be obtained simply by the following substitution,
\bea
\label{subst}
F_{m5} \quad \longrightarrow \quad
F_{m5}+2J_m\da(z) \; ,
\eea
which makes the (modified) transformations non-singular
\emph{when the boundary condition $B_m^{(+)}=J_m$ is taken
into account}. (From our analysis \cite{my1} of the Mirabelli and Peskin
model, we know that we need this substitution, and \emph{not}
$F_{m5}\;\rightarrow\;F_{m5}+2B_m^{(+)}\da(z)$. The reason
is that the supersymmetry variations of $B_m^{(+)}$ and $J_m$
are \emph{different}, which plays a role when the supersymmetry
algebra is calculated.)

Explicitly, the modified supersymmetry transformations are
\bea
\da_\Eta^\prime \psi_{m1} &=& \da_\Eta \psi_{m1}
-\frac{8i}{2\sx}(\si_m{}^n+\da_m^n)\eta_1 J_n e_\hf^5\da(z)
\nn\\[5pt]
\da_\Eta^\prime \psi_{m2} &=& \da_\Eta \psi_{m2} 
-\frac{8i}{2\sx}(\si_m{}^n+\da_m^n)\eta_2 J_n e_\hf^5\da(z)
\nn\\[5pt]
\da_\Eta^\prime \psi_{51} &=& \da_\Eta \psi_{51}
-\frac{8}{2\sx}\si^n\etabar_2 J_n\da(z)
+\frac{8i}{2\sx}e_\hf^m (\si_m{}^n+\da_m^n)\eta_1 J_n\da(z)
\nn\\[5pt]
\da_\Eta^\prime \psi_{52} &=& \da_\Eta \psi_{52}
+\frac{8}{2\sx}\si^n\etabar_1 J_n\da(z)
+\frac{8i}{2\sx}e_\hf^m (\si_m{}^n+\da_m^n)\eta_2 J_n\da(z)
-4\eta_2^{(+)}\da(z) \; .
\nn\\
\eea
It remains to see whether these are the correct modifications
for a particular model. What we can check at the moment
is the closure of the supersymmetry algebra on the bosonic
fields.

Let us consider the commutator of the two (modified)
supersymmetry transformations on $B_5$. We find,
\bea
[\da_\Xi^\prime, \da_\Eta^\prime] B_5 
&=& i\frac{\sx}{2}(\eta_1\da_\Xi^\prime\psi_{52}-\eta_2\da_\Xi^\prime\psi_{51})
\nn\\[5pt]
&=& [\da_\Xi, \da_\Eta] B_5
+\Big\{
2i\sx(\eta_2^{(+)}\xi_1-\eta_1\xi_2^{(+)})\da(z)
\nn\\[5pt]
&&+4i(\eta_1\si^n\xibar_1+\eta_2\si^n\xibar_2)J_n\da(z)
+4(\eta_2\xi_1-\eta_1\xi_2)e_\hf^n J_n\da(z)
+h.c.
\Big\} \; .
\nn\\
\eea
From the (original) supersymmetry algebra, Eq.~(\ref{algebra}),
we know that
\bea
[\da_\Xi, \da_\Eta] B_5 &=& \da_v B_5 +\da_u B_5
\nn\\[5pt]
&=& v^n\p_n B_5+v^5\p_5 B_5+B_n\p_5 v^n+B_5\p_5 v^5
+\p_5 u \; ,
\eea
where (see Eq.~(\ref{saparameters}))
\bea
u &=& -v^m B_m -v^5 B_5+u_0
\nn\\[5pt]
v^m &=& 2i(\eta_1\si^m\xibar_1+\eta_2\si^m\xibar_2)
+2e_\hf^m(\eta_2\xi_1-\eta_1\xi_2)+h.c.
\nn\\[5pt]
u_0 &=&  -i\sx(\eta_2\xi_1-\eta_1\xi_2)+h.c.
\eea
We can, therefore, write
\bea
[\da_\Xi, \da_\Eta] B_5 =v^n F_{n5}+\p_5 u_0 \; ,
\eea
whereas for the commutator of the modified supersymmetry
transformations we find
\bea
[\da_\Xi^\prime, \da_\Eta^\prime] B_5 =v^n [F_{n5}+2J_n\da(z)]
+[\p_5 u_0-2u_0^{(+)}\da(z)] \; .
\eea
This is one explicit check of the fact that the commutator 
of the modified supersymmetry transformations closes
onto the modified gauge transformations. (It is, actually,
obvious. The commutator of two \emph{non-singular} transformations
must be \emph{non-singular}!) 

But in order that the algebra of the modified gauge
transformations close
\emph{without} the use of the boundary conditions, we should
correct our modified general coordinate transformations, 
Eq.~(\ref{modgct}), by replacing there $B_m^{(+)}$ with $J_m$
(and similarly for other odd fields).
Then,
\bea
\da_v^\prime B_m = \da_v B_m -2 v^5 J_m\da(z) \; .
\eea
The modified transformation is, therefore, non-singular 
\emph{only when the boundary condition $B_m^{(+)}=J_m$ is taken
into account}. If the natural boundary condition for $B_m$ is
$B_m^{(+)}=0$, then no modification is necessary.

\section{Brane-localized matter}
\label{sec-sg12}
In this section we argue that the preservation of the
bulk $U(1)$ gauge invariance is \emph{necessary} if the
brane-localized matter is to provide a non-zero boundary
condition for the bulk $B_m$ field. We show that a
seemingly gauge non-invariant boundary condition 
$B_m^{(+)}=J_m$ can in fact be gauge invariant, if the
brane fields transform appropriately under the bulk 
$U(1)$ transformation. We also find that a similar boundary
condition exists in the Horava and Witten model.

\subsection{Preserving the bulk $U(1)$ gauge invariance}
Let us discuss the addition of the brane-localized matter
a little bit further. As in Section \ref{sec-modsusy}, 
all we will use is that
the boundary condition for $B_m$ is modified to
$B_m^{(+)}=J_m$, where $J_m$ is a composite of
the brane-localized fields.

We found that we need to make a modification of our gauge
transformations by making them non-singular. The modified
transformations coincide with those of the boundary picture
\emph{exactly}, both in the bulk and on the brane/boundary.
Therefore, all the conclusions of the boundary picture discussion
hold in the orbifold picture as well. In particular, if the 
$U(1)$ gauge invariance is broken, then 
the closure of the supersymmetry algebra requires
(because of Eq.~\ref{uBm}) 
the $B_m^{(+)}=0$ boundary condition. 
But now this boundary condition is
\emph{inconsistent} with the natural boundary condition
$B_m^{(+)}=J_m$. The only way out is to \emph{preserve}
the $U(1)$ gauge invariance.

We can try the following approach.
In parallel with the modification of the supersymmetry 
transformations, we make the same substitution (\ref{subst}),
\bea
\label{newF}
F_{m5}  \longrightarrow 
F_{m5}^\prime = F_{m5}+2J_m\da(z)  \; ,
\eea
in the bulk action.
The modified action then has $B_m^{(+)}=J_m$ as its natural
boundary condition. Indeed, the $B_M$ equation of motion is
now
\bea
D_M F^{MK}-\frac{3}{6\sx}\eps^{MNPQK}F_{MN}F_{PQ} =0
\eea
(omitting the 2-Fermi terms), where $F_{m5}$ should be
replaced by $F_{m5}^\prime$. The cancellation of the singular terms
requires $F_{m5}^\prime$ to be non-singular, which determines
the jump of $B_m$ across the brane. The parity assignment
($B_m$ is odd) then implies the boundary condition
$B_m^{(+)}=J_m$.

Under the modified $U(1)$ gauge transformation, Eq.~(\ref{modU1}),
\bea
\da_u^\prime B_5 =\p_5 u -2u^{(+)}\da(z) \; ,
\eea
the original $F_{m5}$ is not invariant, whereas the modified 
$F_{m5}^\prime$ can be made invariant, if we choose a special
transformation for $J_m$,
\bea
\da_u^\prime J_m =\p_m u^{(+)} \quad \Rightarrow \quad
\da_u^\prime F_{m5}^\prime =0 \; .
\eea
Since the gauge transformation for $B_m$ is unmodified,
$\da_u^\prime B_m=\p_m u$, its restriction on the brane is
\bea
\da_u^\prime B_m^{(+)} =\p_m u^{(+)} \quad \text{on}\; \Si \; ,
\eea
which means that the boundary condition $B_m^{(+)}=J_m$
is now \emph{gauge invariant}! 

But the modified bulk-plus-brane action is \emph{not} 
yet gauge invariant. Its variation under the modified 
$U(1)$ gauge transformation is still given by
Eq.~(\ref{CScontr2}). It appears that there is no way
to cancel it, at least \emph{without} the use of the
boundary condition for $B_m$. (Note that adding any term
with $B_m$ in the brane action would now break the construction, 
giving a different boundary condition for $B_m$.)
But if we use the $B_m^{(+)}=J_m$ boundary condition, 
we can write Eq.~(\ref{CScontr2}) as follows,
\bea
\label{CScontr3}
\da^\prime_u S = \int d^5x e_4 \da(z) \left\{
u^{(+)}\frac{2}{6\sx}\eps^{mnpq}J_{mn}J_{pq} \right\} \; ,
\eea
where $J_{mn}=\p_m J_n-\p_n J_m$. This, in principle, can
be canceled by the variation of the brane action (excluding
terms which combine into $F_{m5}^\prime = F_{m5}+2J_m\da(z)$).

For example, suppose that the matter fields on the
brane include a scalar $\phi$ and two vectors, $A_m$ and $C_m$. 
Take their transformations under the bulk $U(1)$ to be as follows,
\bea
\da_u^\prime \phi=u^{(+)}, \qquad
\da_u^\prime A_m=\p_m u^{(+)}, \qquad
\da_u^\prime C_m=0 \; .
\eea
Let the brane action (before coupling)
contain $\phi$ and $A_m$ only via
$A_{mn}\equiv \p_m A_n -\p_n A_m$ and 
$\mc{D}_m\phi=\p_m\phi-A_m$, so that it is gauge invariant.
Now couple the brane fields to the bulk supergravity in
the way described above, taking 
\bea
J_m=\p_m\phi+C_m \; ,
\eea
which has the correct gauge transformation, 
$\da_u^\prime J_m=\p_m u^{(+)}$. The resulting action is
not yet gauge invariant, because Eq.~(\ref{CScontr3}) gives
\bea
\da^\prime_u S = \int d^5x e_4 \da(z) \left\{
u^{(+)}\frac{2}{6\sx}\eps^{mnpq}C_{mn}C_{pq} \right\} \; ,
\eea
where $C_{mn}\equiv \p_m C_n -\p_n C_m$. But it is now
easy to make it gauge invariant by adding a term of
the form
$\phi\,\eps^{mnpq}C_{mn}C_{pq}$
to the brane action.

We conclude, therefore, that adding brane-localized matter
can help restore the invariance of the bulk-plus-brane
action (and of the associated with it natural boundary conditions) 
under the bulk $U(1)$ gauge transformation. The preservation
of this invariance is \emph{necessary} for supersymmetry as
we argued based on the closure of the supersymmetry algebra.

We also would like to emphasize that it appears to be impossible 
to maintain the gauge invariance \emph{without} the use of the 
$B_m$ boundary condition.

\subsection{Modified Bianchi identity}
The modified field strength, $F_{MN}^\prime$, which we introduced
satisfies the following modified Bianchi identity,
\bea
(d F^\prime)_{5nk} = \frac{1}{3}
(\p_5 F_{nk}^\prime+\p_n F_{k5}^\prime-\p_k F_{n5}^\prime)
= \frac{2}{3}J_{nk}\da(z) \; ,
\eea
where $(d F)_{MNK}\equiv \p_{[M}F_{NK]}$ and 
$J_{mn}\equiv\p_m J_n-\p_n J_m$.
This is not a surprise, 
since our construction in the previous subsection is
analogous to the construction used by Horava and Witten \cite{hw}.
But while they came to the modification of the bulk field strength 
in order to preserve the 
\emph{brane-localized gauge invariance}, 
we need it to preserve the 
\emph{bulk gauge invariance}.
And our modification is \emph{forced} on us by the supersymmetry algebra!

Note that Horava and Witten write their boundary condition,
Eq.~(2.20) in Ref.~\cite{hw}, in a form which appears to be manifestly
gauge invariant (under both the brane \emph{and} the bulk gauge
invariances). The form is analogous to our $F_{mn}^{(+)}=J_{mn}$.
However, this relation is \emph{derivative} from the basic
boundary condition $B_m^{(+)}=J_m$. There is a similar
not manifestly gauge invariant boundary condition in the
Horava and Witten model.

The boundary condition $B_m^{(+)}=J_m$
follows from our (modified) action as a natural
boundary condition, since now $F_{m5}^\prime$ must be non-singular
to avoid uncanceled singularities in the $B_M$ equation of motion.
It is exactly what Horava and Witten say after Eq.~(2.18) in 
Ref.~\cite{hw}.
Their Eq.~(2.13) then implies the boundary condition
$C_{ABC}^{(+)}\sim \om_{ABC}$, from which their Eq.~(2.20)
follows. This boundary condition can be made invariant under
the brane gauge invariance, provided that $C_{ABC}$
transforms under the brane gauge transformation
like $\om_{ABC}$ in their Eq.~(2.14). 
However, it is unclear how to achieve
the \emph{bulk gauge invariance} 
($\da C_{IJK}=\p_{[I}\Lambda_{JK]}$ 
with \emph{arbitrary} $\Lambda_{IJ}$)
of this boundary condition.

The boundary condition $C_{ABC}^{(+)}\sim \om_{ABC}$ was
also found in Refs.~\cite{moss1, moss2} (although there it
was not derived as a natural boundary condition following
from the action, but simply imposed for consistency).
The transformation of $C_{ABC}$ necessary to preserve
the brane gauge invariance is provided there, but the preservation
of the bulk gauge invariance 
is not discussed.

\newpage
\section{Summary and Conclusions}
\label{sec-sg13}
In this paper we constructed a bulk-plus-boundary action with
the five-dimensional gauged (on-shell) supergravity in the bulk
which is supersymmetric upon the use of
the minimum set of boundary 
conditions dictated by the supersymmetry algebra. In a general
case when the supersymmetry parameter $\Eta_i$ is restricted
by $\eta_2=\al\eta_1$ on the boundary, we found that only the
boundary condition $B_m=0$ have to be used to prove supersymmetry
to second order in fermions. Other boundary conditions following
from the action, $K_{ma}=\la_1 e_{ma}$ and $\psi_{m2}=\al\psi_{m1}$
on $\p\mc{M}$, are not needed in the proof of supersymmetry of the
action.

The necessary ingredient of our boundary action is the 
Gibbons-Hawking-like $Y$-term presented in Eq.~(\ref{Yterm}),
\bea
Y=K+(\psi_{m1}\si^{mn}\psi_{n2}+h.c.)+\ga^{mn} F_{m\hf} B_n
\eea
(where we used $e_5^\hf F^{m5}=\ga^{mn}F_{n\hf}$ as follows
from Eq.~(\ref{simpleFm5})).
It includes
\begin{itemize}
\item[1)]
the standard Gibbons-Hawking term (the trace of
the extrinsic curvature) which allows the derivation of
the boundary condition $K_{ma}=\la_1 e_{ma}$ as a natural 
boundary condition corresponding to the variation $\da e_{ma}$
\cite{barth};
\item[2)]
a fermionic term, which leads to the derivation
of $\psi_{m2}=\al\psi_{m1}$ as a natural boundary condition
for $\da\psi_{m1}$;
\item[3)]
another bosonic term, which lets us derive the $B_m=0$
boundary condition as a natural boundary condition.
\end{itemize}
We argued that the $Y$-term can be derived most easily from
the fact that it must match onto the brane-localized singularities
of the bulk Lagrangian. (This can be used to derive appropriate
additions to the $Y$-term when higher order Fermi terms are
considered.)

In the transition to the orbifold picture, the $Y$-term disappears.
The rest of the boundary action becomes (after the multiplication by 2)
the brane action with which supersymmetry of the 
bulk-plus-brane action in the orbifold
picture can once again be proven using only the minimum set of the
boundary conditions.
We found, however, that one also has to choose unconventional
$\ep(z)$ assignments for the odd fields and parameters,
Eqs.~(\ref{neweps}),
\bea
\eta_2=\ep(z)\eta_2^{(+)}, \quad 
\psi_{m2}=\frac{1}{\ep(z)} \psi_{m2}^{(+)}, \quad
K_{ma}=\frac{1}{\ep(z)} K_{ma}^{(+)}, \quad
q_3=\frac{1}{\ep(z)} q_3^{(+)} \; .
\eea
and use the property $\ep(z)^{-2}\da(z)=-\da(z)$.

The reason for such $\ep(z)$ assignments is unclear. (Perhaps,
the explanation can come from a smooth realization of the
supersymmetric Randall-Sundrum scenario.) We can only observe
that together with the Eq.~(\ref{neweps3}), $u=\ep(z)u^{(+)}$
(where $u$ is the odd parameter of the $U(1)$ gauge transformation),
there is an indication that odd parameters of local transformations
come with $\ep(z)$, whereas other fields and parameters come with
$1/\ep(z)$. (There is a slight problem with such a conclusion,
because the equation (\ref{neweps2}) and the discussion in 
Section \ref{secDifEp} seem to indicate that
\bea
B_m=\frac{1}{\ep(z)} B_m^{(+)}, \quad
e_{5a}=\ep(z)e_{5a}^{(+)} \; .
\eea
But the evidence provided for these assignments is not on a
very firm footing.)

Another important conclusion of this work is that in the
orbifold picture \emph{all} local transformations have to
be modified by the addition of brane-localized terms.
The modifications must be such that the modified transformations
become \emph{non-singular} on the brane when the natural
boundary conditions (encoded in the action) are used.
This is the reason both for the modification of $\da_\Eta \psi_{52}$ in
Refs.~\cite{abn, bb1} and for the ``modification of the Bianchi
identity'' in the Horava-Witten model \cite{hw}.

We also note that our results (concerning the boundary picture)
are in agreement with the recent work of Moss \cite{moss1,moss2},
who did a similar analysis for the eleven-dimensional supergravity.
The use of the fermionic boundary condition there is necessary,
according to our discussion, precisely because higher order
fermionic terms are considered. Our approach to the $Y$-term
can be employed there to derive the boundary condition for
$C_{ABC}$ from the bulk-plus-boundary action (instead of just
postulating it for consistency).

Finally, it would be interesting to see how the analysis
presented here for the case of on-shell supergravity
can be done for \emph{off-shell} supergravity of Zucker \cite{zu4}.
One question this could answer is whether the boundary 
condition $B_m=0$ is, actually, an ``auxiliary boundary condition''
(see Section \ref{secaux}) similar to $\Phi=0$ in the
Mirabelli and Peskin model (in the absence of brane-localized
matter) \cite{my1}.

\acknowledgments
I would like to thank
Jonathan Bagger for his interest in this work, 
many helpful discussions and critical reading of
the manuscript. 
This work was supported in part by the National Science Foundation, 
grant NSF-PHY-0401513.

\appendix
\newpage
\section{Conventions}
\label{app-sg0}
We follow conventions of Ref.~\cite{bb1}. Our indices are
\be
\begin{array}{c@{\hspace{15pt}}c@{\hspace{15pt}}c@{\hspace{15pt}}c}
M,N,P,Q,K & \text{curved space} & M=\{m,5\} & 
m=\{0,1,2,3\}\\[1mm]
A,B,C,D,E & \text{tangent space} & A=\{a,\hat5\} & 
a=\{\hat0,\hat1,\hat2,\hat3\}\\[1mm]
i,j & SU(2) & i=\{1,2\}.& {}
\end{array}
\ee
We denote the determinant of an $n$-bein by $e_n$:
$e_5={\rm det}e_M^A$, $e_4={\rm det}e_m^a$.
We use the f\"unfbein $e_M^A$ to relate the two types of indices,
e.g.
\be
g_{MN} = e_M^A e_N^B \eta_{AB}, \quad
\eps^{MNPQK} = e_A^M e_B^N e_C^P e_D^Q e_E^K \eps^{ABCDE} \, .
\ee
It also defines the torsion-free connection,
\be
\om(e)_{MAB}=\frac{1}{2}e_A^N e_B^K(C_{MNK}+C_{NMK}-C_{KMN}) \; ,
\ee
where $C_{MNK}=e_{MC}(\p_N e_K^C-\p_K e_N^C)$. The covariant
derivative is defined to act as follows,
\bea
D(\om)_M\Psi_N^A=\p_M\Psi_N^A+\om_{MC}{}^A\Psi_N^C
-\Ga(\om)_{MN}^K\Psi_K^A+\frac{1}{4}\om_{MBC}\Ga^{BC}\Psi_N^A \; ,
\eea
where the spinor indices on $\Psi$ and $\Ga^{BC}$ are implicit.
The Christoffel connection is made dependent by imposing 
$D_M e_N^A=0$, which implies
\be
\Ga_{MN}^K=\om_{MN}{}^K+e_A^K\p_M e_N^A \, .
\ee
The curvature tensor is defined by
\be
R_{MNAB}=\p_{M}\om_{NAB}-\p_N\om_{MAB}
+\om_{NA}{}^C\om_{MCB} - \om_{MA}{}^C\om_{NCB} \, ,
\ee
and the scalar curvature is 
$R = e^{MA}R_{MA} = e^{MA}e^{NB}R_{MNAB}$.

The gamma matrices obey the following relations,
\bea
\{\Ga^A, \Ga^B\}=-2\eta^{AB}, \qquad \Ga^{ABCDE}=-\eps^{ABCDE} \nn\\[5pt]
\Ga^{ABCD}=\eps^{ABCDE}\Ga_E, \qquad 
\Ga^{ABC}=\frac{1}{2}\eps^{ABCDE}\Ga_{DE} \; ,
\eea
where $\Ga^{A_1\dots A_n}$ are antisymmetrized 
with ``strength one'', e.g. $\Ga^{AB}=\frac{1}{2}(\Ga^A\Ga^B-\Ga^B\Ga^A)$.
The metric and the Levi-Civita tensor are determined by
\bea
\eta_{AB} = {\rm diag} (-  +  +  +  +), \qquad
\eps^{\hat0\hat1\hat2\hat3\hat5} = +1, \quad
\eps^{abcd\hf} = \eps^{abcd} \; .
\eea
In reduction to the two-component notation \cite{wb} we use
the following representation of the gamma matrices,
\bea
&&\Ga^a = \bpm 0&\si^a \\ \sibar^a&0 \epm, \quad
\Ga^{ab}=2\bpm \si^{ab} & 0 \\ 0 & \sibar^{ab} \epm, \quad
\Ga^{abc}=i\eps^{abcd}\bpm 0 & \si_d \\ -\sibar_d & 0 \epm 
\\[5pt]
&&\Ga^\hf = \bpm -i&0 \\ 0&i \epm, \quad
\Ga^{a\hf}=i\bpm 0 & \si^a \\ -\sibar^a & 0 \epm, \quad
\Ga^{ab\hf}=2i\bpm -\si^{ab} & 0 \\ 0 & \sibar^{ab} \epm .
\eea
A four-component Dirac spinor $\Psi$, its Dirac conjugate $\Psibar$
and its Majorana conjugate $\Psitil$ are written in terms
of two-component spinors $\psi_1$ and $\psi_2$ as follows,
\bea
\Psi = \binom{\psi_1}{\psibar_2}, \quad
\Psibar = (\psi_2 , \; \psibar_1), \quad
\Psitil = (-\psi_1 , \; \psibar_2).
\eea
A symplectic Majorana spinor $\Psi_i$ satisfies $\Psitil^i = \Psibar_i$,
where index $i$ can be raised and lowered with an antisymmetric
tensor $\ep_{ij}$. We use the following representation,
\bea
\Psi_1 = -\Psi^2 = \binom{ \psi_1}{\psibar_2}, \quad
\Psi_2 =  \Psi^1 = \binom{-\psi_2}{\psibar_1} .
\eea
The following identities are satisfied,
\bea
\wt\Psi\dvec\Ga\Eta=\wt\Eta\bvec\Ga\Psi, \qquad
\wt\Psi^i\dvec\Ga\Eta_i=-\wt\Eta^i\bvec\Ga\Psi_i, \qquad
Q_i{}^j\wt\Psi^i\dvec\Ga\Eta_j=Q_i{}^j\wt\Eta^i\bvec\Ga\Psi_j \; ,
\eea
where $\dvec\Ga = \Ga^{A_1}\Ga^{A_2}\dots\Ga^{A_n}$,
$\bvec\Ga = \Ga^{A_n}\dots\Ga^{A_2}\Ga^{A_1}$ and
\bea
\Qij = i\vec q \cdot \vec\si=i\left(
\bma
q_3 & q_1-iq_2\\
q_1+iq_2 & -q_3
\ema\right) .
\eea
Also, for arbitrary symplectic Majorana spinors we have
\bea
&&i\Psitil^i\dvec\Ga\Eta_i=
i\Psibar_1\dvec\Ga\Eta_1 +h.c.
\nn\\[5pt]
&& i\Psitil^i\dvec\Ga\Qij\Eta_j=
-q_3\Psibar_1\dvec\Ga\Eta_1 -q_{12}\Psibar_2\dvec\Ga\Eta_1 +h.c.\;,
\eea
where $q_{12}=q_1+iq_2$. These identities allow 
a straightforward reduction of the action and supersymmetry transformations
to the two-component expressions. The following set of expressions
is especially helpful,
\bea
&\Psibar\Eta=\psi_2\eta_1+\psibar_2\etabar_2, \qquad
\Psibar\Ga^\hf\Eta=-i(\psi_2\eta_1-\psibar_1\etabar_2)& \nn\\
&\Psibar\Ga^a\Eta=\psi_2\si^a\etabar_2+\psibar_1\sibar^a\eta_1, \qquad
\Psibar\Ga^{a\hf}\Eta=i(\psi_2\si^a\etabar_2-\psibar_1\sibar^a\eta_1)& \nn\\
&\Psibar\Ga^{ab}\Eta=2(\psi_2\si^{ab}\eta_1
                       +\psibar_1\sibar^{ab}\etabar_2), \qquad
\Psibar\Ga^{ab\hf}\Eta=-2i(\psi_2\si^{ab}\eta_1
                           -\psibar_1\sibar^{ab}\etabar_2)& .
\eea
Finally,
\bea
\Psibar\Ga^{abcd}\Eta=\eps^{abcd}\Psibar\Ga^\hf\Eta, \qquad
\Psibar\Ga^{abc\hf}\Eta=-\eps^{abcd}\Psibar\Ga_d\Eta .
\eea

\newpage
\section{Gibbons-Hawking boundary term}
\label{app-sg1}
We define the extrinsic curvature as
\footnote{For a detailed discussion of the extrinsic
curvature see Refs.~\cite{mtw,york3,wald}.
}
\bea
K_{MN}=P_M{}^K P_N{}^L D_K n_L \; ,
\eea
where $n^M$ is the (outward pointing) unit vector 
normal to the boundary $\p\mc{M}$,
and 
\bea
P_M{}^N=\da_M^N-n_M n^N
\eea
is a projector onto the boundary, $P_M{}^N n_N=0$.
The trace of the extrinsic curvature is
\bea
K=g^{MN}K_{MN} \; .
\eea
One can show \cite{thesis} that its general variation gives
\bea
\da K=K_{MN}e_A^M \da e^{NA}
+n_M(e^{MA}e^{NB}\da\om_{NAB})
+P_M{}^K D_K(e_A^L P_L{}^M n_N\da e^{NA}) \; .
\eea
The last term is a total tangential derivative which vanishes
upon integrating over $\p\mc{M}$. Since
\bea
\da \int_\mc{M} \left( -\frac{1}{2}R \right)
=\int_\mc{M} \left( R_A^M-\frac{1}{2}R e_A^M \right)\da e_M^A
+\int_{\p\mc{M}}\left( -n_M e^{MA}e^{NB}\da\om_{NAB} \right) \; ,
\eea
we see that the Einstein-Hilbert action with 
the Gibbons-Hawking boundary term,
\bea
S_{EH+GH} =
-\frac{1}{2}\int_\mc{M} R 
+\int_{\p\mc{M}} K \; ,
\eea
under the general variation gives
\bea
\da S_{EH+GH}
=\int_\mc{M} \left( R_A^M-\frac{1}{2}R e_A^M \right)\da e_M^A
+\int_{\p\mc{M}}\left( K_{MN}-K P_{MN} \right)e_A^N\da e^{MA} \; .
\eea
The Gibbons-Hawking term makes only the variation of the metric
(vielbein $e^{MA}$) appear in the boundary term 
of the general variation of the total action.
This improves the variational principle, allowing both the use of the
Dirichlet boundary conditions for the metric,
\bea
\da e_M^A=0 \quad \text{on}\;\p\mc{M} \; ,
\eea
and the derivation of the ``natural'' (generalized Neumann) boundary 
conditions,
\bea
K_{MN}-K P_{MN}=S_{MN} \quad \text{on}\;\p\mc{M} \; ,
\eea
where $S_{MN}$ represents a contribution from a boundary action.

\section{Why we choose $n_5=-e_5^\hf$}
\label{app-sg2}
The Stokes's theorem states (see, e.g., Ref.~\cite{wald})
\bea
\int_\mc{M} d^5x e_5 (D_M K^M) =
\int_\mc{M} d^5x \p_M(e_5 K^M) =
\int_{\p\mc{M}} d^4x e_4^{\rm{ind}} (n_M K^M) \; ,
\eea
where (denoting by $g_{mn}^{\rm{ind}}$ the induced 
four-dimensional metric on $\p\mc{M}$)
\bea
e_5 &=& \det e_{MA} =\sqrt{|\det g_{MN}|} \\[5pt]
e_4^{\rm{ind}} &=& \det e_{ma}^{\rm{ind}} 
=\sqrt{|\det g_{mn}^{\rm{ind}}|} \; ,
\eea
and $n_M$ 1) is orthogonal to $\p\mc{M}$;
2) has the unit norm, $g^{MN}n_M n_N=1$;
3) is outward pointing. With our description of $\p\mc{M}$
as a hypersurface $x^5=\rm{const}$, the first condition
implies that only $n_5\neq 0$, the second says
\bea
n_5=\pm \frac{1}{\sqrt{g^{55}} } \; ,
\eea
and the third has to do with choosing one of the two
signs.

In our gauge ($e_m^\hf=0$), we have $g^{55}=e_\hf^5 e_\hf^5 $
and $e_5^\hf e_\hf^5=1$, thus 
\bea
n_5=\pm e_5^\hf \; .
\eea
In this gauge we also have $g_{mn}^{\rm{ind}}=e_m^a e_{na}$
and, therefore, we can choose the induced vierbein as
$e_{ma}^{\rm{ind}}=e_{ma}$ and obtain $e_5=e_4 e_5^\hf$.

Let us assume that our $\mc{M}$ is a strip 
$x^5\equiv z\in [z_1, z_2]$ and $\Si$ denotes an $x^5=\rm{const}$
hypersurface.
The Stokes's theorem can then be written as
\bea
\int_{\Si} d^4x \int_{z_1}^{z_2} dz \p_5(e_4 e_5^\hf K^5)
=\int_{\p\mc{M}} d^4x e_4 (n_5 K^5) \; ,
\eea
and, therefore,
\bea
(e_5^\hf K^5)_{|z_2}-(e_5^\hf K^5)_{|z_1} =
(n_5 K^5)_{|z_2}+(n_5 K^5)_{|z_1} \; .
\eea
This means that for the Stokes's theorem to hold (that is
for $n_M$ to be ``outward pointing''), we should choose
\bea
n_5=-e_5^\hf \;\; \text{at} \;\; z_1 \qquad \text{and} \qquad
n_5=+e_5^\hf \;\; \text{at} \;\; z_2 \; .
\eea
This choice coincides with the intuitive one when $e_5^\hf>0$.

When $\mc{M}=\mc{M}_+=\mathbb{R}^{1,3}\times[0, +\infty)$,
the outward pointing $n_M$ at $z=0$
has $n_5=-e_5^\hf$. With this choice,
the boundary conditions we obtain on $\p\mc{M}$ coincide
with the boundary conditions ``on the positive side of the brane''
(that is ``at $z=+0$'') and thus directly correspond to 
the boundary conditions in Refs.~\cite{bb1} and \cite{bb2}.

\section{Our gauge}
\label{app-sg3}
Our gauge choice is $e_m^\hf=0$. Thus,
\bea
e_m^\hf=0, \qquad e_a^5=0, 
\qquad e_5^a\neq 0, \qquad e_\hf^m\neq 0  \; .
\eea
Since $e_A^M$ is the inverse to $e_M^A$,
\bea
e_M^A e_A^N =\da_M^N, \qquad e_A^M e_M^B =\da_A^B  \; ,
\eea
in this gauge we have
\bea
e_m^a e_a^n=\da_m^n, \qquad 
e_a^m e_m^b=\da_a^b, \qquad
e_5^\hf e_\hf^5 =1, \qquad
e_\hf^m =-e_5^a e_a^m e_\hf^5  \; .
\eea

\subsection{Metric tensor}
For the metric tensor $g_{MN}$ we obtain
\bea
g_{mn}=\ga_{mn}, \qquad
g_{m5}=g_{5m}=N_m, \qquad
g_{55}= \ga^{mn}N_m N_n+N^2 \; ,
\eea
where we defined
\bea
\ga_{mn}\equiv e_m^a e_{na}, \qquad
N_m\equiv e_5^a e_{ma}, \qquad
N=e_5^\hf \; ,
\eea
and $\ga^{mn}$ is the inverse to $\ga_{mn}$,
\bea
\ga^{mn}\equiv e^{ma}e_a^n, \quad \ga_{mk}\ga^{kn}=\da_m^n \; .
\eea
Defining 
\bea
N^m \equiv \ga^{mn}N_n= e_5^a e_a^m=-N e_\hf^m \; ,
\eea
the inverse five dimensional metric tensor $g^{MN}$
can be written as
\bea
g^{mn}=\ga^{mn}+N^{-2}N^m N^n, \qquad
g^{m5}=g^{5m}=-N^{-2}N^m, \qquad
g^{55}=N^{-2} \; .
\eea

\subsection{Field strength}
Components of $F^{MN}=g^{MK}g^{NL}F_{KL}$ for the field
strength $F_{MN}=\p_M B_N-\p_N B_M$ are
\bea
F^{mn} &=& g^{mk}g^{nl}F_{kl}+(g^{mk}g^{n5}-g^{nk}g^{m5})F_{m5} 
\nn\\[5pt]
F^{m5} &=& g^{mn}g^{k5}F_{nk}+(g^{mn}g^{55}-g^{m5}g^{n5})F_{n5} \; .
\eea
We can write this in a more convenient form,
\bea
\label{simpleFm5}
F^{mn}=\ga^{mk}\ga^{nl}F_{kl}
+\ga^{mk}e_\hf^n F_{k\hf}-\ga^{nk}e_\hf^m F_{k\hf}, \qquad
F^{m5}=\ga^{mn}e_\hf^5 F_{n\hf} \; ,
\eea
where we defined
\bea
F_{m\hf}\equiv e_\hf^5 F_{m5}+e_\hf^n F_{mn} \; .
\eea

\subsection{Spin connection}
We use the following spin connection,
\bea
\om_{MAB}=e_A^N e_B^K\om_{MNK}, \qquad
\om_{MNK}=\frac{1}{2}(C_{MNK}+C_{NMK}-C_{KMN}) \; ,
\eea
where
\bea
C_{MNK}=e_{MC}(\p_N e_K^C -\p_K e_N^C) \; .
\eea
We find that in our gauge
\bea
&&C_{mnk} = \wh C_{mnk}, \quad
C_{mn5} = u_{mn} \nn\\[5pt]
&&C_{5mn} = N^k\wh C_{kmn}, \quad
C_{5n5} = N^k u_{kn}+N\p_n N \; ,
\eea
where
\bea
\wh C_{mnk} \equiv e_{mc}(\p_n e_k^c -\p_k e_n^c), \qquad
u_{mn} \equiv e_{mc}(\p_n e_5^c -\p_5 e_n^c) \; .
\eea
The spin connection coefficients are given by
\bea
\om_{mab} &=& \frac{1}{2}e_a^n e_b^k
(\wh C_{mnk}+\wh C_{nmk}-\wh C_{kmn}) \\
\om_{5ab} &=& \frac{1}{2}N^n\wh C_{nab}
-\frac{1}{2}(u_{ab}-u_{ba}) \\
\om_{ma\hf} &=& -\frac{1}{2}N^{-1}N^k(\wh C_{mak}+\wh C_{amk})
+\frac{1}{2}N^{-1}(u_{am}+u_{ma}) \\
\om_{5a\hf} &=& -\frac{1}{2}N^{-1}N^n N^k\wh C_{nak}
+\frac{1}{2}N^{-1}N^k(u_{ka}+u_{ak})
+e_a^n\p_n N \; .
\eea
We see that in our gauge there are no $\p_5 N_m$ in
any of $\om_{MAB}$, which means that the spin connection
coefficients are \emph{non-singular}, i.e. contain no $\da(z)$,
in the orbifold picture!

\subsection{Extrinsic curvature}
Our (outward pointing) unit vector normal to the boundary 
$\p\mc{M}$ is
\bea
n_M =(0_m, -N), \qquad n^M=(N^{-1}N^m, -N^{-1}) \; .
\eea
This gives the following projector onto the boundary,
\bea
P^{MN}=\left(\bma \ga^{mn} & 0 \\[5pt] 0 & 0 \ema\right) \quad
P_M{}^N=\left(\bma \da_m^n & 0 \\[5pt] N^n & 0 \ema\right) \quad
P_{MN}=\left(\bma \ga_{mn} & N_m \\[5pt] N_n & N^k N_k \ema\right) \; ,
\eea
and the extrinsic curvature,
\bea
K^{MN}=\left(\bma K^{mn} & 0 \\[5pt] 0 & 0 \ema\right) \quad
K_{MN}=\left(\bma K_{mn} & N^k K_{mk} \\[5pt]
N^k K_{kn} & N^k N^l K_{kl} \ema\right) \; ,
\eea
where
\bea
K_{mn}=-\Ga_{mn}^5 n_5, \qquad K^{mn}=\ga^{mk}\ga^{nl}K_{kl} \; .
\eea
The trace of the extrinsic curvature is
\bea
K\equiv g_{MN}K^{MN}=\ga_{mn}K^{mn}=\ga^{mn}K_{mn}
=-\ga^{mn}\Ga_{mn}^5 n_5 \; .
\eea
Using the relation between the Christoffel symbols and the
spin connection,
\bea
D_M e_N^A \equiv \p_M e_N^A+\om_{MC}{}^A e_N^C
-\Ga_{MN}^K e_K^A =0 \; ,
\eea
we find
\bea
\Ga_{mn}^5 = \om_{mn}{}^5+e_A^5\p_m e_n^A
&=& e_n^a\om_{ma}{}^\hf e_\hf^5 \nn\\
&&+ \, e_n^a\om_{ma}{}^b e_b^5
+e_n^\hf\om_{m\hf}{}^b e_b^5
+e_a^5\p_m e_n^a +e_\hf^5\p_m e_n^\hf \; .
\eea
But all the terms in the second line vanish in our gauge, so
\bea
\Ga_{mn}^5 = e_n^a\om_{ma}{}^\hf e_\hf^5 \; .
\eea
Therefore, we obtain
\bea
K_{ma}\equiv e_a^n K_{mn} =\om_{ma\hf}, \qquad
K=e^{ma}K_{ma} \; .
\eea
This gives a geometrical meaning to the spin connection
coefficient $\om_{ma\hf}$ .

\subsection{Advantages of the $e_m^\hf=0$ gauge}
The following properties are unique to our gauge, $e_m^\hf=0$.
(Another simple gauge, $e_5^a = 0$, frequently used in the Kaluza-Klein
reductions,\footnote{
See, e.g., the paper by Chamseddine and Nicolai in Ref.~\cite{hist}.}
\emph{does not} enjoy these properties.)
\begin{enumerate}
\item
$e_m^a$ is an induced vierbein on a slice $x^5=\text{const}$.
\item
$\om_{mab}$ is a spin connection for $e_m^a$.
\item
$\wh D_m e_n^a = 
\p_m e_n^a+\om_{mc}{}^a e_n^c-\Ga_{mn}^k e_k^a=0$, where
$\om_{mab}$ and $\Ga_{mn}^k$ are elements of $\om_{MAB}$
and $\Ga_{MN}^K$.
\item
There is a simple relation between the extrinsic curvature and
a spin connection coefficient:
$K_{ma}=\om_{ma\hf}$ .
\item
There are no $\da(z)$-terms in any of the spin connection 
coefficients $\om_{MAB}$.
\item
The compensating local Lorentz rotation 
(with the parameter $\om^{a\hf}$; see below)
leaves the supersymmetry transformation of $e_m^a$ unchanged.
\end{enumerate}

\section{Supersymmetry transformations}
\label{app-sg4}
In our gauge and in the two-component spinor notation, the 
supersymmetry transformations of Eqs.~(\ref{5Dsusytr}--\ref{5Dsusytr3})
(dropping the 3-Fermi terms in the $\da\Psi_{Mi}$)
can be written as follows,
\bea
\da_\Eta e_m^a &=&
-i(\psi_{m1}\si^a\etabar_1+\psi_{m2}\si^a\etabar_2)+h.c. 
\nn\\[5pt]
\da_\Eta e_5^a &=&
-i(\psi_{51}\si^a\etabar_1+\psi_{52}\si^a\etabar_2)+h.c. 
\nn\\[5pt]
\da_\Eta e_m^\hf &=& -\psi_{m2}\eta_1+\psi_{m1}\eta_2 +h.c. 
\nn\\[5pt]
\da_\Eta e_5^\hf &=& -\psi_{52}\eta_1+\psi_{51}\eta_2 +h.c. 
\nn\\[5pt]
\da_\Eta B_m &=& i\frac{\sx}{2}(\psi_{m2}\eta_1-\psi_{m1}\eta_2)+h.c. 
\nn\\[5pt]
\da_\Eta B_5 &=& i\frac{\sx}{2}(\psi_{52}\eta_1-\psi_{51}\eta_2)+h.c. 
\eea
\bea
\da_\Eta\psi_{m1} &=& 2\wh D_m\eta_1+i\om_{ma\hf}\si^a\etabar_2
+i\la\si_m(q_{12}^\ast\etabar_1+q_3\etabar_2)
-i\sx\la(q_3\eta_1-q_{12}^\ast\eta_2)B_m 
\nn\\[5pt]
&&+\frac{1}{2\sx}\Big\{
-4i(\si_m{}^n+\da_m^n)\eta_1 
(e_\hf^5 F_{n5}+e_\hf^k F_{nk})
+[i\eps_m{}^{nkl}\si_l+4\da_m^n\si^k]\etabar_2 F_{nk} \Big\}
\nn\\[5pt]
\da_\Eta\psi_{m2} &=& 2\wh D_m\eta_2-i\om_{ma\hf}\si^a\etabar_1
+i\la\si_m(q_3\etabar_1-q_{12}\etabar_2)
+i\sx\la(q_3\eta_2+q_{12}\eta_1)B_m 
\nn\\[5pt]
&&+\frac{1}{2\sx}\Big\{
-4i(\si_m{}^n+\da_m^n)\eta_2
(e_\hf^5 F_{n5}+e_\hf^k F_{nk})
-[i\eps_m{}^{nkl}\si_l+4\da_m^n\si^k]\etabar_1 F_{nk} \Big\}
\nn\\[5pt]
\da_\Eta\psi_{51} &=& 2\wh D_5\eta_1+i\om_{5a\hf}\si^a\etabar_2
+\la(e_5^\hf-i\sx B_5)(q_3\eta_1-q_{12}^\ast\eta_2)
\nn\\[5pt]
&&+\frac{1}{2\sx}\Big\{
-4(\si^n\etabar_2-i e_\hf^n\eta_1)F_{n5}
-4i e_{5a}\si^{an}\eta_1(e_\hf^5 F_{n5}+e_\hf^k F_{nk})
\nn\\[2pt]
&&\hspace{160pt}
+(-2i e_5^\hf \si^{nk}\eta_1
+i e_{5a}\eps^{ankl}\si_l\etabar_2)F_{nk} \Big\}
\nn\\[5pt]
\da_\Eta\psi_{52} &=& 2\wh D_5\eta_2-i\om_{5a\hf}\si^a\etabar_1
-\la(e_5^\hf-i\sx B_5)(q_3\eta_2+q_{12}\eta_1)
\nn\\[5pt]
&&+\frac{1}{2\sx}\Big\{
+4(\si^n\etabar_1+i e_\hf^n\eta_2)F_{n5}
-4i e_{5a}\si^{an}\eta_2(e_\hf^5 F_{n5}+e_\hf^k F_{nk})
\nn\\[2pt]
&&\hspace{160pt}
+(-2i e_5^\hf \si^{nk}\eta_2
-i e_{5a}\eps^{ankl}\si_l\etabar_1)F_{nk} \Big\} \; ,
\eea
where
\bea
\wh D_M\eta =\p_M\eta+\frac{1}{2}\om_{Mab}\si^{ab}\eta \; .
\eea
Note that \emph{all} $e_5^a$ and $e_\hf^m$ have been explicitly
separated out (thus, $\si^m=e_a^m\si^a$ and so on).

However, we have to modify the supersymmetry transformations by 
a compensating Lorentz transformation ($\om_{AB}=-\om_{BA}$),
\bea
\da_\om e_M^A = e_M^B\om_B{}^A, \qquad
\da_\om \Psi_{Mi} = \frac{1}{4}\om_{AB}\Ga^{AB}\Psi_{Mi} \; ,
\eea
in order to stay in our $e_m^\hf=0$ gauge. Using
only off-diagonal coefficient $\om^{a\hf}$,
\bea
\da_\om e_m^a =-e_{m\hf}\om^{a\hf}, \quad
\da_\om e_5^a=-e_{5\hf}\om^{a\hf}, \quad
\da_\om e_m^\hf=e_{mc}\om^{c\hf}, \quad
\da_\om e_5^\hf=e_{5c}\om^{c\hf} \; .
\eea
We define the modified supersymmetry transformations by
\bea
\da_\Eta^\prime e_M^A =\da_\Eta e_M^A + \da_\om e_M^A \; .
\eea
We want the supersymmetry variation to preserve $e_m^\hf=0$,
\bea
\da_\Eta^\prime e_m^\hf=\da_\Eta e_m^\hf +e_{mc}\om^{c\hf} =0 \; ,
\eea
which fixes 
\bea
\om^{a\hf} =-e^{ma}\da_\Eta e_m^\hf
=e^{ma}(\psi_{m2}\eta_1-\psi_{m1}\eta_2)+h.c.
\eea
The modified supersymmetry transformations, therefore, are
\bea
\da_\Eta^\prime e_m^a &=& \da_\Eta e_m^a \\
\da_\Eta^\prime e_5^a &=& \da_\Eta e_5^a +e_{5\hf}e^{ma}\da_\Eta e_m^\hf \\
\da_\Eta^\prime e_m^\hf &=& 0\\
\da_\Eta^\prime e_5^\hf &=& \da_\Eta e_5^\hf -e_5^a e_a^m\da_\Eta e_m^\hf \; .
\eea
Note that the supersymmetry variation of the $e_m^a$ stays the same,
which is one of the advantages of our gauge.

The gravitino supersymmetry transformations also get modified, but
only in the 3-Fermi terms (since $\om^{a\hf}$ is 2-Fermi) which we omit.

\section{Bulk Lagrangian}
\label{app-sg5}
The fermionic part of the bulk supergravity Lagrangian is
\bea
\mc{L}_{5F} &=&
\frac{i}{2}\Psitil_M^i \Ga^{MNK} D_N \Psi_{Ki} 
-i\frac{\sx}{8}F^{MN}\Psitil_M^i\Psi_{Ni}
-i\frac{\sx}{16}F_{MN}\Psitil_P^i\Ga^{MNPQ}\Psi_{Qi} \nn\\
&&+i\frac{3}{4}\la\Qij\Psitil_M^i\Ga^{MN}\Psi_{Nj}
-i\frac{\sx}{4}\la\Qij\Psitil_M^i\Ga^{MNK}\Psi_{Kj}B_N \; .
\eea
In the two-component spinor notation it becomes
\bea
\label{L5Fcomp}
\mc{L}_{5F} &=&
\frac{1}{2}\eps^{mnkl}(\psibar_{m2}\sibar_l D_n\psi_{k2}
                     +\psibar_{m1}\sibar_l D_n\psi_{k1})
+(\psi_{m1}\si^{mn}D_\hf \psi_{n2}
 -\psi_{m2}\si^{mn}D_\hf \psi_{n1}) 
\nn\\[5pt]
&&
+(\psi_{m2}\si^{mn}D_n\psi_{\hf 1}
 -\psi_{\hf 1}\si^{mn}D_m\psi_{n2})
+(\psi_{\hf 2}\si^{mn}D_m\psi_{n1}
 -\psi_{m1}\si^{mn}D_n\psi_{\hf 2}) 
\nn\\[5pt]
&&
-i\frac{\sx}{4}
\Big[\ga^{mk}\ga^{nl}F_{kl}(\psi_{m2}\psi_{n1})
    +\ga^{mk}F_{k\hf}(\psi_{m2}\psi_{\hf 1}
                     -\psi_{m1}\psi_{\hf 2})
\Big] 
\nn\\[5pt]
&&
-\frac{\sx}{8}\eps^{mnpq}
\Big[ F_{pq}(\psi_{m2}\psi_{n1})
   -i F_{pq}(\psi_{m2}\si_n\psibar_{\hf 2}
            +\psi_{m1}\si_n\psibar_{\hf 1}) 
\nn\\[5pt]
&&\hspace{150pt}
   -i F_{m\hf}(\psi_{p2}\si_n\psibar_{q2}
              +\psi_{p1}\si_n\psibar_{q1})
\Big]
\nn\\[5pt]
&&
-\frac{3}{2}\la
\Big\{
   q_3\big[2\psi_{m1}\si^{mn}\psi_{n2}
           +i(\psi_{m2}\si^m\psibar_{\hf 2}
             -\psi_{m1}\si^m\psibar_{\hf 1}) 
       \big]
\nn\\[5pt]
&&\hspace{30pt}
  +q_{12}\big[(\psi_{m1}\si^{mn}\psi_{n1}
                -\psibar_{m2}\sibar^{mn}\psibar_{n2})
             +i(\psi_{m1}\si^m\psibar_{\hf 2}
               +\psibar_{m2}\sibar^m\psi_{\hf 1})
          \big]
\Big\} 
\nn\\[5pt]
&&
+\frac{\sx}{4}\la
\Big\{
   q_3\big[ i B_n\eps^{mnkl}(\psi_{m2}\si_l\psibar_{k2}
                            -\psi_{m1}\si_l\psibar_{k1})
           +4i B_\hf (\psi_{m1}\si^{mn}\psi_{n2}) 
\nn\\[5pt]
&&\hspace{150pt}
           -4i B_n (\psi_{m2}\si^{mn}\psi_{\hf 1}
                   +\psi_{m1}\si^{mn}\psi_{\hf 2})
       \big] 
\nn\\[5pt]
&&\hspace{40pt}
  +q_{12}\big[ 2i B_n\eps^{mnkl}(\psi_{m1}\si_l\psibar_{k2})
              +2i B_\hf (\psi_{m1}\si^{mn}\psi_{n1}
                        +\psibar_{m2}\sibar^{mn}\psibar_{n2}) 
\nn\\[5pt]
&&\hspace{180pt}
              -4i B_n (\psi_{m1}\si^{mn}\psi_{\hf 1}
                      +\psibar_{m2}\sibar^{mn}\psibar_{\hf 2})
          \big]
\Big\}
\nn\\[5pt]
&&
+ h.c.
\eea
Note that $e_5^a$ does not appear at all, whereas $e_\hf^m$
appears \emph{only} in the following combinations,
\bea
&& B_\hf=e_\hf^5 B_5+e_\hf^m B_m, \qquad
F_{m\hf}=e_\hf^5 F_{m5}+e_\hf^n F_{mn} \nn\\[5pt]
&& D_\hf=e_\hf^5 D_5+e_\hf^m D_m, \qquad
\psi_{\hf 1,2}=e_\hf^5 \psi_{5 1,2}+e_\hf^m \psi_{m 1,2} \; .
\eea
In particular, $\si^m=e_a^m\si^a$ and 
$\eps^{mnkl}=e_a^m e_b^n e_c^k e_d^l \eps^{abcd}$.
The derivatives can be further decomposed as follows,
\bea
D_M\psi_1 &=& \wh D_M\psi_1+\frac{i}{2}\om_{Ma\hf}\psibar_2 \nn\\
D_M\psi_2 &=& \wh D_M\psi_2-\frac{i}{2}\om_{Ma\hf}\psibar_1 \; .
\eea
Note that $D_M e_N^A=0$, but $\wh D_M e_N^A \neq 0$.



\begin{thebibliography}{99}

\bibitem{hw}
  P.~Horava and E.~Witten,
  ``Eleven-Dimensional Supergravity on a Manifold with Boundary,''
  Nucl.\ Phys.\ B {\bf 475}, 94 (1996)
  [arXiv:hep-th/9603142].

\bibitem{rs1}
  L.~Randall and R.~Sundrum,
  ``A large mass hierarchy from a small extra dimension,''
  Phys.\ Rev.\ Lett.\  {\bf 83}, 3370 (1999)
  [arXiv:hep-ph/9905221].

\bibitem{rs2}
  L.~Randall and R.~Sundrum,
  ``An alternative to compactification,''
  Phys.\ Rev.\ Lett.\  {\bf 83}, 4690 (1999)
  [arXiv:hep-th/9906064].

\bibitem{abn}
R.~Altendorfer, J.~Bagger and D.~Nemeschansky,
``Supersymmetric Randall-Sundrum scenario,''
Phys.\ Rev.\ D {\bf 63}, 125025 (2001)
[hep-th/0003117].

\bibitem{gp1}
T.~Gherghetta and A.~Pomarol,
``Bulk fields and supersymmetry in a slice of AdS,''
Nucl.\ Phys.\ B {\bf 586}, 141 (2000)
[hep-ph/0003129].

\bibitem{flp1}
A.~Falkowski, Z.~Lalak and S.~Pokorski,
``Supersymmetrizing branes with bulk in five-dimensional supergravity,''
Phys.\ Lett.\ B {\bf 491}, 172 (2000)
[hep-th/0004093].

\bibitem{bb1}
  J.~Bagger and D.~V.~Belyaev,
  ``Supersymmetric branes with (almost) arbitrary tensions,''
  Phys.\ Rev.\ D {\bf 67}, 025004 (2003)
  [arXiv:hep-th/0206024].

\bibitem{bb2}
  J.~Bagger and D.~Belyaev,
  ``Twisting warped supergravity,''
  JHEP {\bf 0306}, 013 (2003)
  [arXiv:hep-th/0306063].


\bibitem{gibh}
  G.~W.~Gibbons and S.~W.~Hawking,
  ``Action Integrals And Partition Functions In Quantum Gravity,''
  Phys.\ Rev.\ D {\bf 15}, 2752 (1977).

\bibitem{york1}
J.~W.~York, Jr.; 
``Role of conformal three-geometry in the dynamics of gravitation,''
Phys.\ Rev.\ Lett.\  {\bf 28}, 1082 (1972).

\bibitem{york2}
J.~W.~York, Jr.; 
``Boundary terms in the action principles of general relativity,''
Foundations of Physics, {\bf 16}, 249 (1986). 

\bibitem{barth}
  N.~H.~Barth,
  ``The Fourth Order Gravitational Action For Manifolds With Boundaries,''
  Class.\ Quant.\ Grav.\  {\bf 2}, 497 (1985).

\bibitem{my1}
D.~V.~Belyaev,
``Boundary conditions in the Mirabelli and Peskin model,''
arXiv:hep-th/0509171.


\bibitem{mp}
  E.~A.~Mirabelli and M.~E.~Peskin,
  ``Transmission of supersymmetry breaking from a 4-dimensional boundary,''
  Phys.\ Rev.\ D {\bf 58}, 065002 (1998)
  [arXiv:hep-th/9712214].

\bibitem{thesis}
D.~V.~Belyaev, ``Five-Dimensional Supergravity on a Manifold
with Boundary,'' Ph.D. thesis.


\bibitem{bkvp}
  E.~Bergshoeff, R.~Kallosh and A.~Van Proeyen,
  ``Supersymmetry in singular spaces,''
  JHEP~{\bf 0010}, 033 (2000)
  [arXiv:hep-th/0007044].

\bibitem{pvn}
  P.~Van Nieuwenhuizen,
  ``Supergravity,''
  Phys.\ Rept.\  {\bf 68}, 189 (1981).




\bibitem{zu4}
  M.~Zucker,
  ``Off-shell supergravity in five-dimensions and supersymmetric brane world
  scenarios,''
  Fortsch.\ Phys.\  {\bf 51}, 899 (2003).

\bibitem{hist}
E.~Cremmer, 
``Supergravities In 5 Dimensions,''
in {\it Superspace and supergravity}, S.W.~Hawking and
M.~Rocek eds., Cambridge University Press, 1981, pp.267--282;\newline
A.~H.~Chamseddine and H.~Nicolai,
``Coupling The SO(2) Supergravity Through Dimensional Reduction,''
Phys.\ Lett.\ B {\bf 96}, 89 (1980);
\newline
R.~D'Auria, E.~Maina, T.~Regge and P.~Fre,
``Geometrical First Order Supergravity In Five Space-Time Dimensions,''
Annals Phys.\  {\bf 135}, 237 (1981);\newline
M.~Gunaydin, G.~Sierra and P.~K.~Townsend,
``The Geometry Of N=2 Maxwell-Einstein Supergravity And Jordan Algebras,''
Nucl.\ Phys.\ B {\bf 242}, 244 (1984);
``Gauging The D = 5 Maxwell-Einstein Supergravity Theories: 
More On Jordan Algebras,''
Nucl.\ Phys.\ B {\bf 253}, 573 (1985).


\bibitem{dive1}
  P.~Di Vecchia, B.~Durhuus, P.~Olesen and J.~L.~Petersen,
  ``Fermionic Strings With Boundary Terms,''
  Nucl.\ Phys.\ B {\bf 207}, 77 (1982).

\bibitem{dive2}
  P.~Di Vecchia, B.~Durhuus, P.~Olesen and J.~L.~Petersen,
  ``Fermionic Strings With Boundary Terms. 2. The O(2) String,''
  Nucl.\ Phys.\ B {\bf 217}, 395 (1983).

\bibitem{lumo}
  H.~Luckock and I.~Moss,
  ``The Quantum Geometry Of Random Surfaces And Spinning Membranes,''
  Class.\ Quant.\ Grav.\  {\bf 6}, 1993 (1989).



\bibitem{gero1}
  G.~von Gersdorff, L.~Pilo, M.~Quiros, A.~Riotto and V.~Sanz,
  ``Fermions and supersymmetry breaking in the interval,''
  Phys.\ Lett.\ B {\bf 598}, 106 (2004)
  [arXiv:hep-th/0404091].

\bibitem{nieu}
  U.~Lindstrom, M.~Rocek and P.~van Nieuwenhuizen,
  ``Consistent boundary conditions for open strings,''
  Nucl.\ Phys.\ B {\bf 662}, 147 (2003)
  [arXiv:hep-th/0211266];
\newline
  P.~van Nieuwenhuizen and D.~V.~Vassilevich,
  ``Consistent boundary conditions for supergravity,''
  arXiv:hep-th/0507172.


\bibitem{conrad}
  J.~O.~Conrad,
  ``Brane tensions and coupling constants from within M-theory,''
  Phys.\ Lett.\ B {\bf 421}, 119 (1998)
  [arXiv:hep-th/9708031].

\bibitem{bds}
  A.~Bilal, J.~P.~Derendinger and R.~Sauser,
  ``M-theory on S(1)/Z(2): new facts from a careful analysis,''
  Nucl.\ Phys.\ B {\bf 576}, 347 (2000)
  [arXiv:hep-th/9912150].



\bibitem{lama2}
  Z.~Lalak and R.~Matyszkiewicz,
  ``Twisted supergravity and untwisted super-bigravity,''
  Phys.\ Lett.\ B {\bf 562}, 347 (2003)
  [arXiv:hep-th/0303227].

\bibitem{gero2}
  G.~von Gersdorff, L.~Pilo, M.~Quiros, A.~Riotto and V.~Sanz,
  ``Supersymmetry from boundary conditions,''
  Nucl.\ Phys.\ B {\bf 712}, 3 (2005)
  [arXiv:hep-th/0411133].



\bibitem{zu3}
  M.~Zucker,
  ``Supersymmetric brane world scenarios from off-shell supergravity,''
  Phys.\ Rev.\ D {\bf 64}, 024024 (2001)
  [arXiv:hep-th/0009083].


\bibitem{matter1}
  T.~Gherghetta and A.~Riotto,
  ``Gravity-mediated supersymmetry breaking in the brane-world,''
  Nucl.\ Phys.\ B {\bf 623}, 97 (2002)
  [arXiv:hep-th/0110022].

\bibitem{matter2}
  R.~Rattazzi, C.~A.~Scrucca and A.~Strumia,
  ``Brane to brane gravity mediation of supersymmetry breaking,''
  Nucl.\ Phys.\ B {\bf 674}, 171 (2003)
  [arXiv:hep-th/0305184].

\bibitem{matter3}
  G.~A.~Diamandis, B.~C.~Georgalas, P.~Kouroumalou and A.~B.~Lahanas,
  ``On the brane coupling of unified orbifolds with gauge interactions in  the
  bulk,''
  Phys.\ Lett.\ B {\bf 602}, 112 (2004)
  [arXiv:hep-th/0402228].

\bibitem{bb3}
  J.~A.~Bagger and D.~V.~Belyaev,
  ``Brane-localized Goldstone fermions in bulk supergravity,''
  Phys.\ Rev.\ D {\bf 72}, 065007 (2005)
  [arXiv:hep-th/0406126].

\bibitem{matter4}
  A.~Falkowski,
  ``On the one-loop Kaehler potential in five-dimensional brane-world
  supergravity,''
  JHEP {\bf 0505}, 073 (2005)
  [arXiv:hep-th/0502072].


\bibitem{moss1}
  I.~G.~Moss,
  ``Boundary terms for eleven-dimensional supergravity and M-theory,''
  Phys.\ Lett.\ B {\bf 577}, 71 (2003)
  [arXiv:hep-th/0308159].

\bibitem{moss2}
  I.~G.~Moss,
  ``Boundary terms for supergravity and heterotic M-theory,''
  arXiv:hep-th/0403106.


\bibitem{wb}
J.~Wess and J.~Bagger, {\em Supersymmetry and Supergravity},
2nd Edition, Princeton University Press, 1992.

\bibitem{mtw}
C.~W.~Misner, K.~S.~Thorne and J.~A.~Wheeler, 
{\it Gravitation}, San Francisco: W.H. Freeman, 1973.

\bibitem{york3}
J.~W.~York, Jr.; "Kinematics and Dynamics of General Relativity"
in: {\it Sources of Gravitational Radiation}, ed. L.L. Smarr,
Cambridge University Press, 1979, pp.~83-126.

\bibitem{wald}
R.~M.~Wald, {\it General Relativity}, 
Chicago: University of Chicago Press, 1984.



\end{thebibliography}
\end{document}